\documentclass[prd,showkeys,floatfix,twocolumn,amsmath,amssymb,floatfix]{revtex4}
\usepackage{amsmath}
\usepackage{graphicx}
\usepackage{float}
\usepackage{epstopdf}
\usepackage{multirow}
\usepackage{subfigure}
\usepackage{color}
\usepackage[colorlinks,citecolor=blue,anchorcolor=red,menucolor=red,linkcolor=red,filecolor=red,runcolor=red,urlcolor=blue,frenchlinks=red]{hyperref}

\allowdisplaybreaks[3]

\begin{document}
\title{2$P$-wave charmed baryons from QCD sum rules}
%

\author{Hui-Min Yang$^1$}
\email{hmyang@pku.edu.cn}
\author{Hua-Xing Chen$^2$}
\email{hxchen@seu.edu.cn}

\affiliation{
$^1$School of Physics and Center of High Energy Physics, Peking University, Beijing 100871, China \\
$^2$School of Physics, Southeast University, Nanjing 210094, China
}

\begin{abstract}
We conduct an investigation on the $1P$- and $2P$-wave charmed baryons using the methods of QCD sum rules and light-cone sum rules within the framework of heavy quark effective theory. Our results suggest that the $\Lambda_c(2910)^+$, $\Lambda_c(2940)^+$, and $\Xi_c(3123)^+$ can be well interpreted as the $2P$-wave charmed baryons of $J^P=1/2^-$ and $3/2^-$, belonging to the $SU(3)$ flavor $\mathbf{\bar 3}_F$ representation. Moreover, the $\Xi_c(3123)^+$ possesses a partner state characterized by $J^P=1/2^-$, denoted as $\Xi_c(1/2^-,2P)$. Our analysis predicts its mass and width to be $m_{\Xi_c(1/2^-,2P)} - m_{\Xi_c(3123)^+} = -18\pm{7}$~MeV and $\Gamma_{\Xi_c(2P,1/2^-)}=31^{+170}_{-~27}$~MeV, with $m_{\Xi_c(3123)^+} = 3122.9\pm{1.3}$~MeV. We propose to search for it in the $\Xi_c(1/2^-,2P)\to \Sigma_c K$ decay channel.
\end{abstract}
\maketitle
\pagenumbering{arabic}
%
%
%
\section{Introduction}\label{sec:intro}
%

The electromagnetic interaction within the deuterium atom contributes to its spectral fine structure, as supported by several studies~\cite{Triebwasser:1953tgb,Dayhoff:1953zz,Schroder:2001rc,Krutov:2011ch,Meissner:2023voo}. Similarly, the strong interaction within the singly heavy baryon, constituted by a heavy quark and two light quarks orbiting around the relatively motionless heavy quark, results in the splitting of the hadron spectroscopy. Hence, this system serves as an excellent framework for investigating this phenomenon~\cite{Isgur:1978wd,Copley:1979wj,Korner:1994nh,Manohar:2000dt,Bianco:2003vb,Karliner:2008sv,Klempt:2009pi} and has attracted significant attentions from both experimental and theoretical communities.

In recent years, substantial strides have been made in the field of heavy baryons. Through both experimental inquiry and theoretical exploration, all ground-state charmed baryons have been accurately identified~\cite{ParticleDataGroup:2020ssz}. Furthermore, various research collaborations have determined the spin-parity quantum numbers of the lowest-lying orbitally excited charmed baryons, namely, $\Lambda_c(2595)$~\cite{CLEO:1994oxm}, $\Lambda_c(2625)$~\cite{ARGUS:1993vtm}, $\Xi_c(2790)$~\cite{CLEO:2000ibb}, and $\Xi_c(2815)$~\cite{CLEO:1999msf}. Subsequently, several newly discovered excited charmed baryons, such as $\Lambda_c(2765)$~\cite{CLEO:2000mbh}, $\Lambda_c(2860)$~\cite{LHCb:2017jym}, $\Lambda_c(2880)$~\cite{CLEO:2000mbh}, $\Lambda_c(2910)$~\cite{Belle:2022hnm}, $\Lambda_c(2940)$~\cite{LHCb:2017jym}, $\Xi_c(2980)$~\cite{Belle:2006edu}, $\Xi_c(3055)$~\cite{BaBar:2007zjt}, $\Xi_c(3080)$~\cite{Belle:2016tai},  $\Xi_c(3123)$~\cite{BaBar:2007zjt} and others, arising from both $B$-decay and $e^+ e^- \to c \bar c$ scattering processes, have been observed by the BaBar, Belle, and LHCb Collaborations. We extract some experimental measurements as follows:
\begin{itemize}
\item In 2006 the BaBar Collaboration observed two narrow charmed baryons, $\Lambda_c(2880)^+$ and $\Lambda_c(2940)^+$~\cite{BaBar:2006itc}, in the $D^0p$ mass distribution. Particularly, the LHCb Collaboration subsequently confirmed the $\Lambda_c(2940)$~\cite{LHCb:2017jym}, and its decay to $\Sigma_c(2455)^{0,++}\pi^{\pm}$ was also observed by Belle~\cite{Belle:2006xni}. Its mass and width were measured to be:
\begin{eqnarray}
\Lambda_c(2940)^+ &:& M = 2939.8 \pm 5.6\pm 3.8 {\rm~MeV} \, ,
\\ \nonumber      && \Gamma = 17.5 \pm 5.2\pm 5.9{\rm~MeV} \, ;
\end{eqnarray}

\item In 2007 the BaBar Collaboration reported two charm-strange baryons, $\Xi_c(3055)^+$ and $\Xi_c(3123)^+$~\cite{BaBar:2007zjt}, in the $\Lambda_c^+K^-\pi^+$ mass distribution, with the parameters of $\Xi_c(3123)^+$ measured as:
\begin{eqnarray}
\Xi_c(3123)^+ &:& M = 3122.9 \pm 1.3\pm 0.3 {\rm~MeV} \, ,
\\ \nonumber      && \Gamma = 4.4 \pm 3.4\pm 1.7{\rm~MeV} \, ;
\end{eqnarray}

\item In 2022 the Belle Collaboration discovered a new structure in the $\Sigma_c(2455)^{0,++}\pi^\pm$ spectrum with a significance of $4.2 \sigma$~\cite{Belle:2022hnm}. This state is tentatively named $\Lambda_c(2910)^+$. Its mass and width were measured to be:
\begin{eqnarray}
\Lambda_c(2910)^0 &:& M = 2913.8 \pm 5.6\pm 3.8{\rm~MeV} \, ,
\\ \nonumber      && \Gamma = 51.8 \pm 20.0\pm 18.8{\rm~MeV}\, .
\end{eqnarray}
\end{itemize}

The Particle Data Group (PDG) have assigned accurate $J^P$ values to the $\Lambda_c(2860)(3/2^+)$, $\Lambda_c(2880)(5/2^+)$, $\Lambda_c(2940)(3/2^-)$, and $\Xi_c(2980)(1/2^+)$, while the properties of $\Lambda_c(2765)$, $\Lambda_c(2910)$, $\Xi_c(3055)$, $\Xi_c(3080)$, and $\Xi_c(3123)$ remain uncertain. A rough estimation of their possible spin-parity values can be made based on the mass discrepancy between the $\Xi_c$ baryons and their corresponding $\Lambda_c$ baryons. Table~\ref{tab:table1} presents all the $\Lambda_c$ and $\Xi_c$ baryons recorded in PDG, with the final column indicating an approximate mass difference of $200$~MeV between $\Xi_c$ and $\Lambda_c$ baryons. Hence, it's logical to assign $J^P=1/2^+$, $3/2^+$, $5/2^+$, and $3/2^-$ to the $\Lambda_c(2765)$, $\Xi_c(3055)$, $\Xi_c(3080)$, and $\Xi_c(3123)$ baryons, respectively. Considering that the $\Lambda_c(2910)$ recently observed by Belle likely corresponds to a companion state of $\Lambda_c(2940)$, it follows that there may also exist a corresponding missing companion state of $\Xi_c(3123)$.

\begin{table*}[hbtp]
\begin{center}
\renewcommand{\arraystretch}{1.5}
\caption{The experimental information of $\Lambda_c$ and $\Xi_c$ baryons. Their masses and decay widths (in units of MeV) are taken from PDG~\cite{ParticleDataGroup:2020ssz}. The mass differences between the $\Lambda_c$ and $\Xi_c$ baryons are listed in the last column.}
\begin{tabular}{ c c  c  c  c c c  c  c}
\hline\hline
Baryons & ~~~~$J^P$~~~~&~~ Mass~~ & ~~Width~~ & Baryons & ~~~~$J^P$~~~~& ~~Mass~~ & ~~Width~~ & ~~$\Delta M$~~
\\ \hline
$\Lambda_c(2286)^+$ & $1/2^+$ & $2286.46\pm 0.14$ &--& $\Xi_c(2468)^0$ &$1/2^+$& $2470.88^{+0.34}_{-0.80}$&--&$184.42^{+0.37}_{-0.81}$
\\
$\Lambda_c(2595)^+$ & $1/2^-$& $2592.25\pm 0.28$&$2.6\pm0.6$&$\Xi_c(2790)^0$ & $1/2^-$ &$2791.8\pm 3.3$&$10.0\pm 1.1$ &$199.6\pm3.3$
\\
$\Lambda_c(2625)^+$ & $3/2^-$ &$2628.11\pm0.19$&$<0.97$&$\Xi_c(2815)^0$& $3/2^-$& $2819.6\pm1.2$&$2.5\pm0.2$&$191.5\pm1.2$
\\
$\Lambda_c(2765)^+$&$?^?$&$2766.6\pm2.4$&$50$&$\Xi_c(2980)^0$&$1/2^+$&$2968.0\pm2.6$&$20\pm7$&$201.4\pm3.5$
\\
$\Lambda_c(2860)^+$&$3/2^+$&$2856.1^{+2.3}_{-5.9}$&$67.6^{+11.8}_{-21.6}$&$\Xi_c(3055)^+$&$?^?$&$3054.2\pm1.3$&$17\pm13$&$198.1^{+2.6}_{-6.0}$
\\
$\Lambda_c(2880)^+$&$5/2^+$&$2881.53\pm0.35$&$5.8\pm1.1$&$\Xi_c(3080)^0$&$?^?$&$3079.9\pm1.4$&$5.6\pm2.2$&$198.4\pm1.4$
\\
$\Lambda_c(2940)^+$&$3/2^-$&$2939.3^{+1.4}_{-1.5}$&$20^{+6}_{-5}$&$\Xi_c(3123)^+$&$?^?$&$3122.9\pm1.3$&$4\pm4$&$183.6^{+1.9}_{-2.0}$
\\ \hline\hline
\end{tabular}
\label{tab:table1}
\end{center}
\end{table*}

The singly charmed baryon system has attracted numerous experimentalists and theorists to study them. However, it remains a challenging issue to fully comprehend their internal structures. Various theoretical methods and models have been applied in this field, including various quark models~\cite{Isgur:1978xj,Ebert:2005xj,Roberts:2007ni,Zhong:2007gp,Bijker:2020tns,Chen:2018vuc,Chen:2018orb,Wang:2018fjm,Xiao:2020gjo,Ebert:2011kk,Chen:2016iyi,Yang:2018lzg,Lu:2020ivo,Wang:2020gkn,Wang:2019uaj,Luo:2019qkm}, the chiral perturbation theory~\cite{Wise:1992hn,Scherer:2002tk,Cheng:2006dk,Lu:2014ina,Cheng:2015naa,Jiang:2015xqa},the chiral effective field theory~\cite{Meng:2019nzy,Wang:2020dhf,Machleidt:2011zz}, various molecular interpretations~\cite{Huang:2018bed,Liang:2017ejq,Chen:2017xat,Liang:2014eba,An:2017lwg,Debastiani:2017ewu,Zhang:2022pxc}, the Regge trajectory~\cite{Guo:2008he,Ebert:2011kk}, the $^3P_0$ model~\cite{Chen:2007xf,Ye:2017yvl,Lu:2019rtg}, the relativistic flux tube model~\cite{Olson:1991tw,Chen:2014nyo}, QCD sum rules~\cite{Aliev:2018vye,Wang:2020pri,Agaev:2020fut,Azizi:2020tgh,Yu:2021zvl,Azizi:2022dpn,Liu:2007fg,Mao:2015gya,Chen:2017sci}, and the lattice QCD~\cite{Padmanath:2013bla,Burch:2015pka,Perez-Rubio:2015zqb,Padmanath:2017lng}, etc. Their production and decay properties have been extensively studied in Refs.~\cite{Chen:2007xf,Ye:2017yvl,Yao:2018jmc,Guo:2019ytq}, and various reviews~\cite{Crede:2013kia,Chen:2016spr,Liu:2019zoy,Cheng:2021qpd,Chen:2022asf} provide insights into the recent progresses.

The higher states $\Lambda_c(2910)^+$ and $\Lambda_c(2940)^+$ among the $\Lambda_c$ family have sparked considerable interest among theorists due to certain perplexing issues. The quantum number $J^P={3/2}^-$ is favored for $\Lambda_c(2940)^+$ according to the LHCb result, while its mass is roughly $60$~MeV smaller than the expected $\Lambda_c({3/2}^-,2P)$ state in the quark models~\cite{Ebert:2011kk,Chen:2014nyo,Chen:2016iyi}. Furthermore, the mass of its partner state $\Lambda_c({1/2}^-,2P)$ is only slightly lighter than that of $\Lambda_c({3/2}^-,2P)$, suggesting that their masses are expected to exceed $3$~GeV. Alternative interpretations of $\Lambda_c(2940)^+$ could involve the introduction of the $D^*N$ channel contribution, as proposed in Refs.~\cite{Zhang:2022pxc,Luo:2019qkm}. These studies propose that the $\Lambda_c(2940)^+$ is probably the isoscalar $D^*N$ molecule encompassing both spin-$1\over2$ and spin-$3\over2$ structures. However, the higher mass of the $\Lambda_c({1/2}^-)$ state leads to a mass inversion. Consequently, it is crucial to verify the quantum numbers of $\Lambda_c(2910)$ and $\Lambda_c(2940)$ through further experimental measurements and theoretical researches.

In the present study we primarily focus on the analyses of the mass spectral and decay properties of the $\Lambda_c(2910)$, $\Lambda_c(2940)$, and $\Xi_c(3123)$ charmed baryons using the QCD sum rules and light-cone sum rules within the heavy quark effective theory (HQET) framework. We examine their $S$- and $D$-wave decays into the ground-state charmed baryons with light pseudoscalar or vector mesons, and calculate the relevant partial widths. The results obtained further support the identification of $\Lambda_c(2910)$, $\Lambda_c(2940)$, and $\Xi_c(3123)$ as $2P$-wave charmed baryons of the $SU(3)$ flavor $\mathbf{\bar 3}_F$. Furthermore, based on our previous work~\cite{Yang:2022oog}, we designated the $\Lambda_c(2595)$, $\Lambda_c(2625)$, $\Xi_c(2790)$, and $\Xi_c(2815)$ as the $1P$-wave charmed baryons. Considering that the $P$-wave interpolating currents derived from Ref.~\cite{Chen:2015kpa} can couple into both $1P$- and $2P$-wave states, it is reasonable to regard the $\Lambda_c(2910)$, $\Lambda_c(2940)$, and $\Xi_c(3123)$ as the first radial excited states of the $\Lambda_c(2595)$, $\Lambda_c(2625)$, and $\Xi_c(2815)$. Additionally, it is noteworthy that a $2P$-wave charmed baryon, the partner state of the $\Xi_c(3123)$, remains undiscovered. In this paper we shall calculate its mass and decay properties.

This paper is organized as follows. In Sec.~\ref{sec:sumrule} we briefly introduce our notations, and apply the QCD sum rule method to calculate the masses of $\Lambda_c(2910)$, $\Lambda_c(2940)$, and $\Xi_c(3123)$ as $2P$-wave charmed baryons of the $SU(3)$ flavor $\mathbf{\bar 3}_F$. The obtained parameters are further used to study their decay properties through the light-cone sum rule method in Sec.~\ref{sec:decay}. In Sec.~\ref{sec:summary} we discuss the results and conclude this paper.

%
\section{Mass analyses from QCD sum rules}
\label{sec:sumrule}
%

\begin{figure*}[hbtp]
\begin{center}
\scalebox{1}{\includegraphics{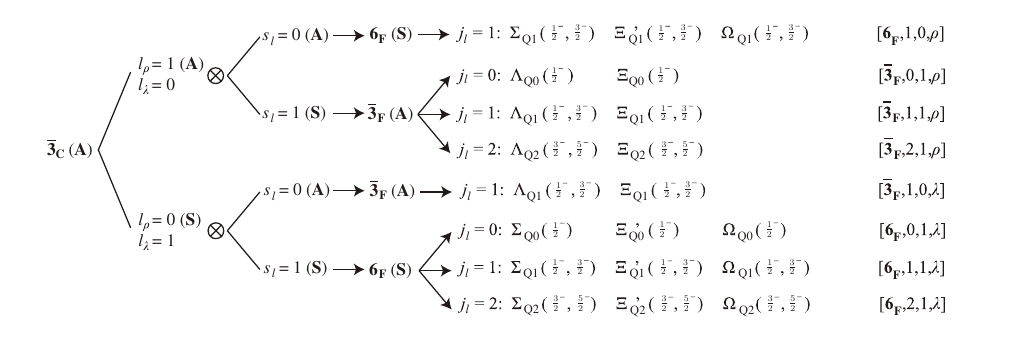}}
\end{center}
\caption{$P$-wave heavy baryons belonging to the $SU(3)$ flavor $\mathbf{\bar 3}_F$ representation.}
\label{fig:pwave}
\end{figure*}

A charmed baryon consists of a heavy charm quark and two light up/down/strange quarks. Its internal structure encompasses various properties, such as the color, flavor, spin, and orbital degrees of freedom. Notably, the orbital excitation of a $P$-wave charmed baryon can occur between the charmed quark and the light quarks, and it can also occur within the two light quarks. The former is known as the $\lambda$-mode excitation, characterized by $l_{\lambda}=1$ and $l_{\rho}=0$, while the latter is termed as the $\rho$-mode excitation with $l_{\lambda}=0$ and $l_{\rho}=1$, as depicted in Fig.~\ref{fig:pwave}. When investigating the singly charmed baryon, it is crucial to consider the properties of the two light up/down/strange quarks:
\begin{itemize}

\item Their color structure is antisymmetric ($\mathbf{\bar 3}_C$).

\item Their flavor structure is either symmetric ($\mathbf{6}_F$) or antisymmetric ($\mathbf{\bar 3}_F$).

\item Their spin structure is either symmetric ($s_l = 1$) or antisymmetric ($s_l = 0$).

\item Their orbital structure is either symmetric ($\lambda$-mode) or antisymmetric ($\rho$-mode).
\end{itemize}

Applying the Pauli principle to the two light quarks, we can categorize the $P$-wave charmed baryons into eight multiplets, denoted as $[F({\rm flavor}), j_l, s_l, \rho/\lambda]$. In this notation $j_l = l_\lambda \otimes l_\rho \otimes s_l$ represents the total angular momentum of the light components. Each multiplet encompasses one or two baryons with the total angular momenta given by $j = j_l \otimes s_c = |j_l \pm 1/2|$. Within the framework of QCD sum rules, we can construct the corresponding interpolating currents $J^{\alpha_1\cdots\alpha_{j-1/2}}_{j,P,F,j_l,s_l,\rho/\lambda}$, which couple to the charmed baryons belonging to the $[F, j_l, s_l, \rho/\lambda]$ multiplet through:
\begin{eqnarray}
\nonumber  \langle 0| J^{\alpha_1\cdots\alpha_{j-{1\over2}}}_{j,P,F,j_l,s_l,\rho/\lambda} |j,P,F,j_l,s_l,\rho/\lambda \rangle = f_{F,j_l,s_l,\rho/\lambda} u^{\alpha_1\cdots\alpha_{j-{1\over2}}}\, ,
\label{eq:coupling}
\end{eqnarray}
with $f_{F,j_l,s_l,\rho/\lambda}$ the decay constant.

As shown in Fig.~\ref{fig:pwave}, there are four multiplets belonging to the $SU(3)$ flavor $\mathbf{\bar 3}_F$ representation. In this work we assume that the $\Lambda_c(2910)$, $\Lambda_c(2940)$, and $\Xi_c(3123)$ are the $2P$-wave charmed baryons of $J^P=1/2^-$ and $J^P=3/2^-$ belonging to the $[\mathbf{\bar 3}_F, 1, 1, \rho]$ doublet. Accordingly, we use the following currents to study them~\cite{Chen:2015kpa}:
\begin{eqnarray}
&& J_{1/2,-,\mathbf{\bar 3_F},1,1,\rho}\label{eq:1/2}\\
\nonumber &=&i\epsilon_{abc}\big([\mathcal{D}_t^\mu q^{aT}]\mathcal{C}\gamma_t^\nu q^b-q^{aT}\mathcal{C}\gamma_t^\nu[\mathcal{D}_t^\mu q^b]\big)\sigma^{\mu\nu}h_v^c\, ,\\
&& J_{3/2,-,\mathbf{\bar 3_F},1,1,\rho}^\alpha \label{eq:32}\\
\nonumber &=&i\epsilon_{abc}\big([\mathcal{D}_t^\mu q^{aT}]\mathcal{C}\gamma_t^\nu q^b-q^{aT}\mathcal{C}\gamma_t^\nu[\mathcal{D}_t^\mu q^b]\big)\\
\nonumber &\times& \big(g_t^{\alpha\mu}\gamma_t^\nu\gamma_5-g_t^{\alpha\nu}\gamma_t^\mu\gamma_5-{1\over3}\gamma_t^\alpha\gamma_t^\mu\gamma_t^\nu\gamma_5+{1\over3}\gamma_t^\alpha\gamma_t^\nu\gamma_t^\mu\gamma_5\big)h_v^c \, .
\end{eqnarray}
Here, $a \cdots c$ represent color indices, $\mathcal{C}$ denotes the charge-conjugation operator, $\mathcal{D}_t^{\mu}=\mathcal{D}^{\mu}-v\cdot \mathcal{D} v^{\mu}$, $\gamma_t^{\nu}=\gamma^{\nu}-v\!\!\!\slash v^{\nu}$, and $g_t^{\rho\beta}=g^{\rho\beta}-v^{\rho} v^{\beta}$. The covariant derivative operator has been explicitly added to these currents.

\subsection{The sum rules at the leading order($m_Q\rightarrow\infty$)}

In this study we assume that the current $J_{1/2,-,\mathbf{\bar 3_F},1,1,\rho}$ can couple into both the $1P$-wave charmed baryons $\Lambda_c(1/2^-,1P)$ and $\Xi_c(1/2^-,1P)$ as well as the $2P$-wave excitations $\Lambda_c(1/2^-,2P)$ and $\Xi_c(1/2^-,2P)$. Similarly, the current $J_{3/2,-,\mathbf{\bar 3_F},1,1,\rho}$ can also couple into both $1P$ and $2P$-wave charmed baryons with $J^P=3/2^-$. If the two light quarks inside a charmed baryon are the up and down quarks, the currents are associated with the $\Lambda_c$; if one of the two light quarks is an strange quark, the currents correspond to the $\Xi_c$. Then, the relation between the state and the relevant interpolating field is:
\begin{eqnarray}
&&\langle 0|J_{1/2,-,\mathbf{\bar 3_F},1,1,\rho}(x)|\Lambda_c(1/2^-,1P/2P)\rangle\\
\nonumber &&~~~~~~~~~~~~~~~~~~~~~~~~~~~~~~~~= f_{\Lambda_c(1/2^-,1P/2P)} u(x)\, ,
\\
&&\langle 0|J_{1/2,-,\mathbf{\bar 3_F},1,1,\rho}(x)|\Xi_c(1/2^-,1P/2P)\rangle\\
\nonumber &&~~~~~~~~~~~~~~~~~~~~~~~~~~~~~~~~= f_{\Xi_c(1/2^-,1P/2P)} u(x)\, ,
\\
&&\langle 0|J_{3/2,-,\mathbf{\bar 3_F},1,1,\rho}(x)|\Lambda_c(3/2^-,1P/2P)\rangle\\ \nonumber 
&&~~~~~~~~~~~~~~~~~~~~~~~~~~~~~~~~= f_{\Lambda_c(3/2^-,1P/2P)} u^\alpha(x)\, ,
\\
&&\langle 0|J_{3/2,-,\mathbf{\bar 3_F},1,1,\rho}(x)|\Xi_c(3/2^-,1P/2P)\rangle\\ \nonumber 
&&~~~~~~~~~~~~~~~~~~~~~~~~~~~~~~~~= f_{\Xi_c(3/2^-,1P/2P)} u^\alpha(x)\, ,
\end{eqnarray}
where $f_{\Lambda_c({1\over2}^-,1P/2P)}$, $f_{\Xi_c({1\over2}^-,1P/2P)}$, $f_{\Lambda_c({3\over2}^-,1P/2P)}$, and $f_{\Xi_c({3\over2}^-,1P/2P)}$ are the decay constants, $u(x)$ and $u^\alpha(x)$ denote the Dirac and Rarita-Schwinger spinors. These currents can be used to construct the two-point correlation function:
\begin{eqnarray}
&&\Pi_{j,P,F,j_l,s_l,\rho/\lambda}^{\alpha_1\cdots \alpha-{j-1/2},\beta_1,\cdots \beta_{j-1/2}}(\omega)\label{eq:two-point}\\
\nonumber &=& i\int d^4 x e^{ikx}\langle 0|T [J_{j,P,F,j_l,s_l,\rho/\lambda}^{\alpha_1\cdots \alpha-{j-1/2}}(x)\bar{J}_{j,P,F,j_l,s_l,\rho/\lambda}^{\beta_1,\cdots \beta_{j-1/2}}(0)]|0\rangle
\\ \nonumber &=& \mathbb{S}[g_t^{\alpha_1\beta_1}\cdots g_t^{\alpha_{j-1/2}\beta_{j-1/2}}]{1+v\!\!\!\slash\over2}\Pi_{j,P,F,j_l,s_l,\rho/\lambda}(\omega)\, ,
\end{eqnarray}
where $\omega=v\cdot k$ is external off-shell energy, and $\mathbb{S}$ denotes symmetrization and subtracting the trace terms in the sets ($\alpha_1\cdots \alpha_{j-1/2}$) and ($\beta_1\cdots \beta_{j-1/2}$).

We employ the four distinct currents for QCD sum rule analyses: $J_{1/2,-,\mathbf{\bar 3_F},1,1,\rho}$, $J_{3/2,-,\mathbf{\bar 3_F},1,1,\rho}$, $J_{1/2,-,\mathbf{\bar 3_F},1,0,\lambda}$, and $J_{3/2,-,\mathbf{\bar 3_F},1,0,\lambda}$. As an illustration of our methodology, we choose the current $J_{1/2,-,\mathbf{\bar 3_F},1,1,\rho}$ to elucidate how to extract the masses of $\Lambda_c({1\over2}^-,1P)$ and $\Lambda_c({1\over2}^-,2P)$. We can write Eq.~(\ref{eq:two-point}) at the hadron level as
\begin{eqnarray}
&&\Pi_{1/2,-,\mathbf{\bar 3_F},1,1,\rho}(\omega,\omega^\prime)\label{eq:hadron}\\
\nonumber &=&{f_{\Lambda_c({1\over2}^-,1P)}^2\over \bar{\Lambda}_{\Lambda_c({1\over2}^-,1P)}-\omega}+{f_{\Lambda_c({1\over2}^-,2P)}^2\over \bar{\Lambda}_{\Lambda_c({1\over2}^-,2P)}-\omega^\prime}+{\rm~higher~~states}\, .
\end{eqnarray}

In this equation $\bar{\Lambda}_{\Lambda_c({1\over2}^-,1P/2P)}$ is defined to be $\bar{\Lambda}_{\Lambda_c({1\over2}^-,1P/2P)}\equiv \lim\limits_{m_c\to \infty}(m_{\Lambda_c(1/2^-,1P/2P)}-m_c)$. To depress the influence of higher-order power, a Borel transformation is performed as
\begin{eqnarray}
 \Pi_{{1/2},-,\mathbf{\bar 3_F},1,1,\rho}&=&f_{\Lambda_c({1\over2}^-,1P)}^2 e^{-\bar{\Lambda}_{\Lambda_c({1\over2}^-,1P)}/T}\\
\nonumber
&+&f_{\Lambda_c({1\over2}^-,2P)}^2e^{-\bar{\Lambda}_{\Lambda_c({1\over2}^-,2P)}/T}\, .
\end{eqnarray}

At the quark-gluon level Eq.~(\ref{eq:two-point}) can be calculated by the method of operator product expansion (OPE). We first insert Eq~(\ref{eq:1/2}) into Eq.~(\ref{eq:two-point}), and then perform the Borel transformation:
\begin{eqnarray}
&&\Pi_{{1/2},-,\mathbf{\bar 3_F},1,1,\rho}(\omega_c^\prime,T)\label{eq:quark}\\
\nonumber &=&\int_0^{\omega_c^\prime} e^{-\omega/T}({3\omega^7\over 35\pi^4}-{\langle g_s^2 G G\rangle\omega^3\over 48\pi^4}) d\omega\\ \nonumber 
&&-{\langle g_s \bar q  \sigma G q\rangle\langle\bar q q\rangle\over 4}-{\langle g_s\bar q  \sigma G q\rangle \langle g_s\bar q \sigma G q\rangle\over 64 T^2}
\\ \nonumber &=& \int_0^{\omega_c} e^{-\omega/T}({3\omega^7\over 35\pi^4}
-{\langle g_s^2 G G\rangle\omega^3\over 48\pi^4}) d\omega\\
\nonumber &&-{\langle  g_s\bar q  \sigma G q\rangle\langle\bar q q\rangle\over 4}
-{\langle g_s\bar q  \sigma G q\rangle \langle g_s\bar q  \sigma G q\rangle\over 64 T^2}\\
\nonumber&&+\int_{\omega_c}^{\omega_c^\prime} e^{-\omega\over T}({3\omega^7\over 35\pi^4}-{\langle g_s^2 G G\rangle\omega^3\over 48\pi^4}) d\omega
\\ \nonumber &=& \Pi_{\Lambda_c({1\over2}^-,1P)}(\omega_c,T)+\Pi_{\Lambda_c({1\over2}^-,2P)}(\omega_c,\omega_c^\prime,T)\, .
\end{eqnarray}
Finally, we differentiate Eq.~(\ref{eq:hadron}) and Eq.~(\ref{eq:quark}) with respect to $(-1/T)$ to obtain $\bar{\Lambda}_{\Lambda_c({1\over2}^-,1P/2P)}$ and $f_{\Lambda_c({1\over2}^-,1P/2P)}$:
\begin{eqnarray}
 \bar{\Lambda}_{\Lambda_c({1\over2}^-,1P)}&=&{{\partial\over \partial(-1/T)}\Pi_{\Lambda_c({1\over2}^-,1P)}\over\Pi_{\Lambda_c({1\over2}^-,1P)}}\, ,
\label{eq:lambdabar121P}\\
 f_{\Lambda_c({1\over2}^-,1P)}&=&\sqrt{\Pi_{\Lambda_c({1\over2}^-,1P)} e^{\bar{\Lambda}_{\Lambda_c({1\over2}^-,1P)}/ T}}\, , \label{eq:f121P}
\\
 \bar{\Lambda}_{\Lambda_c({1\over2}^-,2P)}&=&{{\partial\over \partial(-1/T)}\Pi_{\Lambda_c({1\over2}^-,2P)}\over\Pi_{\Lambda_c({1\over2}^-,2P)}}\, ,\label{eq:lambdabar122P}
\\
 f_{\Lambda_c({1\over2}^-,2P)}&=&\sqrt{\Pi_{\Lambda_c({1\over2}^-,2P)} e^{\bar{\Lambda}_{\Lambda_c({1\over2}^-,2P)}/ T}}\, .\label{eq:f122P}
\end{eqnarray}

In the calculations we work at the renormalization scale of $1$~GeV and use the condensates and other parameters with the following values~\cite{Yang:1993bp,Hwang:1994vp,Ovchinnikov:1988gk,Jamin:2002ev,Ioffe:2002be,Gimenez:2005nt,Colangelo:1998ga}:

\begin{eqnarray}
\nonumber && \langle \bar qq \rangle = - (0.24 \pm 0.01 \mbox{ GeV})^3 \, ,
\\ \nonumber && \langle \bar ss \rangle = (0.8\pm 0.1)\times \langle\bar qq \rangle \, ,
\\ && \langle g_s \bar q \sigma G q \rangle = M_0^2 \times \langle \bar qq \rangle\, ,
\label{eq:condensates}
\\ \nonumber && \langle g_s \bar s \sigma G s \rangle = M_0^2 \times \langle \bar ss \rangle\, ,
\\ \nonumber && M_0^2= 0.8 \mbox{ GeV}^2\, ,
\\ \nonumber && \langle g_s^2GG\rangle =(0.48\pm 0.14) \mbox{ GeV}^4\, ,
\\ \nonumber && m_c = 1.275 \pm 0.035 \mbox{ GeV}\, ,
\\ \nonumber && m_s =128 ^{+12}_{-4}  \mbox{ MeV}\, .
\end{eqnarray}

Eqs.~(\ref{eq:lambdabar121P}-\ref{eq:f122P}) imply that the mass and decay constant depend on three free parameters: the threshold values $\omega_c$ and $\omega_c^\prime$ as well as the Borel mass $T$. There are three criteria to restrict these three parameters: a) the convergence (CVG) of OPE requires that the high-order power corrections are less than $10\%$; b) we require the pole contribution (PC) of OPE to be larger than $15\%$ and $30\%$ with respect to the thresholds $\omega_c$ and $\omega_c^\prime$, respectively; c) the mass dependence on these three parameters is sufficiently stable:
\begin{eqnarray}
{\rm CVG}&\equiv& |{\Pi_{1/2,-,\mathbf{\bar 3_F},1,1,\rho}^{\rm high-order}(T)\over \Pi_{1/2,-,\mathbf{\bar 3_F},1,1,\rho}(\infty,T)}|\leq 10\%\, ,\label{eq:cvg}\\
{\rm PC}&\equiv& {\Pi_{1/2,-,\mathbf{\bar 3_F},1,1,\rho}(\omega_c^\prime,T)\over\Pi_{1/2,-,\mathbf{\bar 3_F},1,1,\rho}(\infty,T)}\ge 30\%\, ,
\label{eq:pc} \\
{\rm PC}^\prime&\equiv& {\Pi_{1/2,-,\mathbf{\bar 3_F},1,1,\rho}(\omega_c,T)\over\Pi_{1/2,-,\mathbf{\bar 3_F},1,1,\rho}(\infty,T)}\ge 15\% \, ,
\label{eq:pcp}
\end{eqnarray}
where $\Pi_{1/2,-,\mathbf{\bar 3_F},1,1,\rho}^{\rm high-order}(T)$ is used to denote the high-order power corrections,
\begin{eqnarray}
&&\Pi_{1/2,-,\mathbf{\bar 3_F},1,1,\rho}^{\rm high-order}(T)\\
\nonumber &=&-{\langle g_s\bar q \sigma G q\rangle\langle\bar q q\rangle\over 4}-{\langle g_s\bar q \sigma G q\rangle \langle \bar q g_s \sigma G q\rangle\over 64 T^2}\, .
\end{eqnarray}

We determine a Borel window $T_{min}\leq T\leq T_{max}$ with fixed values of $\omega_c^\prime$ and $\omega_c$. These parameters, $\omega_c^\prime$ and $\omega_c$, are two independent variables setting at $1.57$~GeV and $1.20$~GeV, respectively, in order to accurately accommodate the masses of $\Lambda_c(1/2^-,1P)$ and $\Lambda_c(1/2^-,2P)$. In this study we derive an interval of $0.260$~GeV$\leq T\leq 0.266$~GeV for $\omega_c^\prime=1.57$~GeV and $\omega_c=1.20$~GeV.

\begin{figure*}[hbt]
\begin{center}
\subfigure[]{
\scalebox{0.46}{\includegraphics{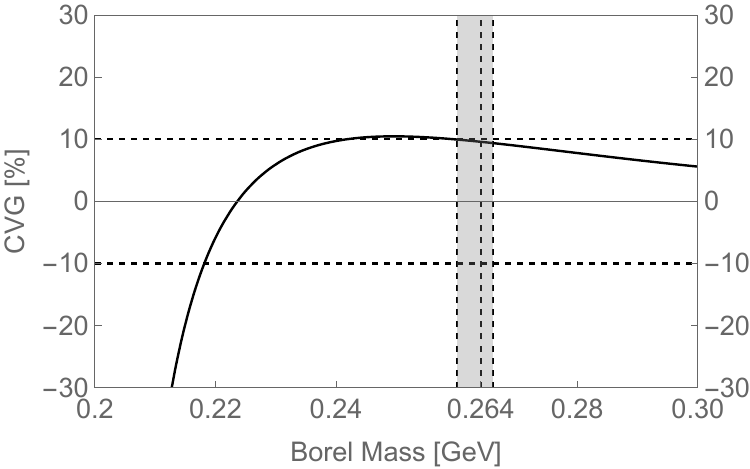}}}~~~~
\subfigure[]{
\scalebox{0.45}{\includegraphics{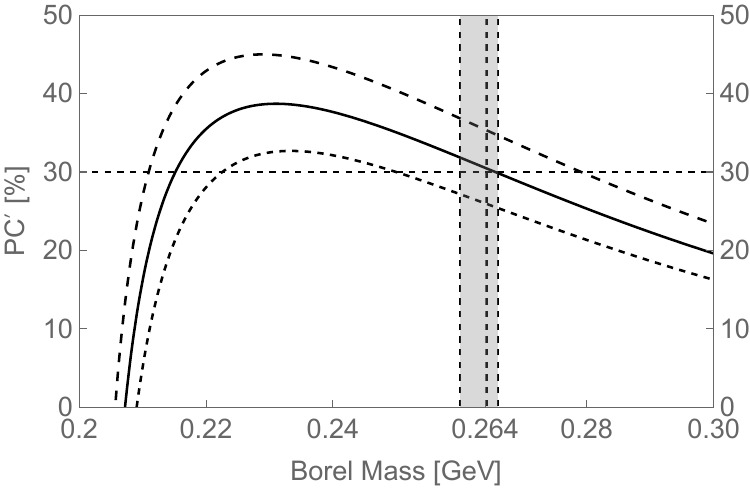}}}~~~~
\subfigure[]{
\scalebox{0.45}{\includegraphics{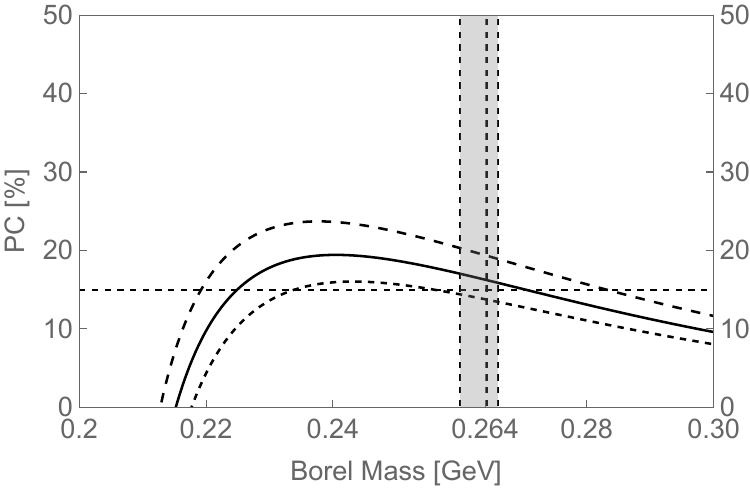}}}
\caption{Variations of CVG (left), PC$^\prime$ (middle), and PC (right), as defined in Eqs.~(\ref{eq:cvg}-\ref{eq:pcp}), with respect to Borel mass $T$, when $J_{1/2,-,\mathbf{\bar 3_F},1,1,\rho}$ is adopted. In the middle panel the short-dashed, solid, and long-dashed curves are obtained by fixing $\omega_c^\prime=1.47$~GeV, $1.57$~GeV, and $1.67$~GeV, respectively. In the right panel the short-dashed, solid, and long-dashed curves are gained by setting $\omega_c=1.10$~GeV, $1.20$~GeV, and $1.30$~GeV, respectively.}
\label{fig:cvgpc}
\end{center}
\end{figure*}

\begin{figure*}[hbt]
\begin{center}
\subfigure[]{
\scalebox{0.5}{\includegraphics{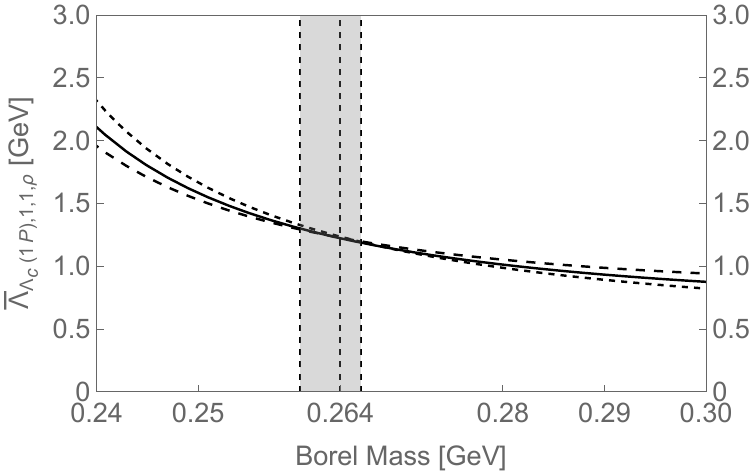}}}~~~~
\subfigure[]{
\scalebox{0.5}{\includegraphics{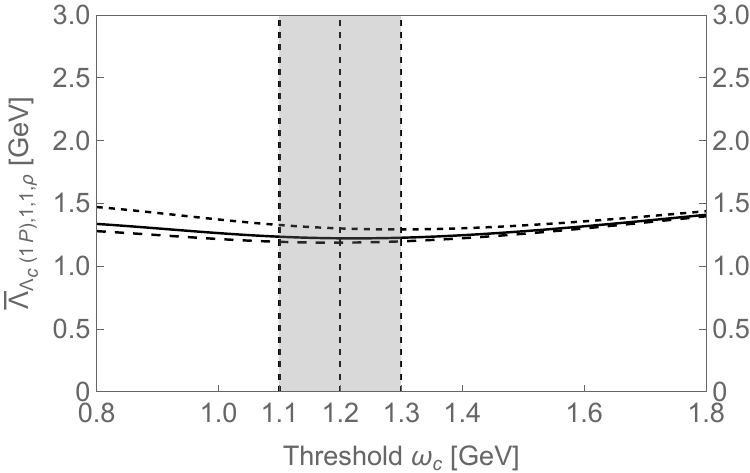}}}~~~~
\\
\subfigure[]{
\scalebox{0.5}{\includegraphics{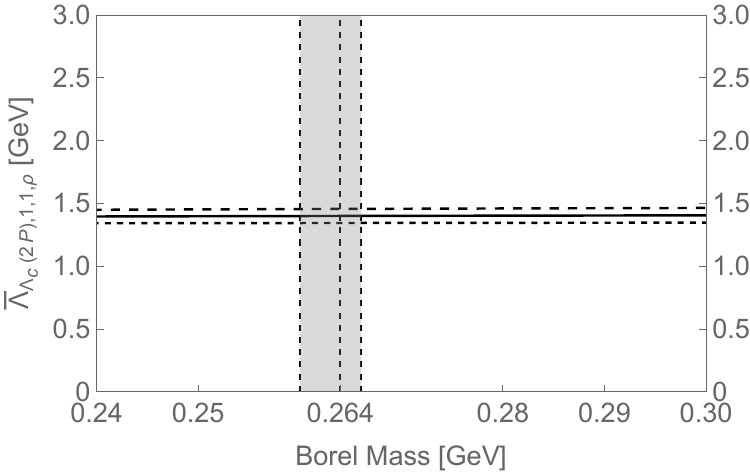}}}~~~~
\subfigure[]{
\scalebox{0.5}{\includegraphics{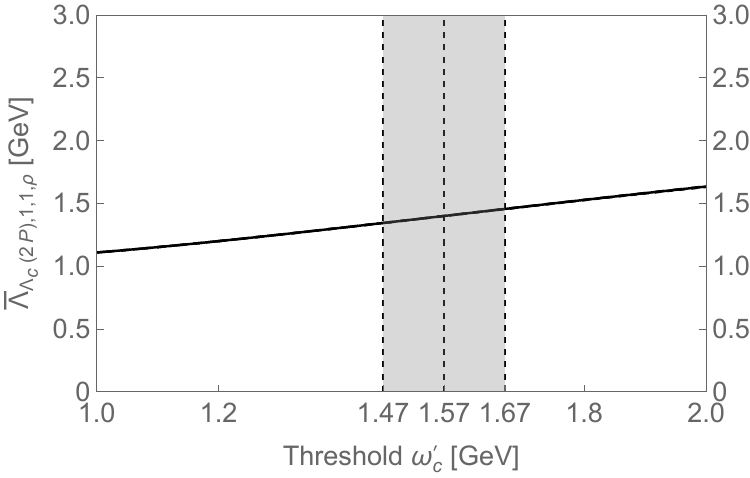}}}~~~~
\caption{Variations of $\bar{\Lambda}_{\Lambda_{c}(1P),1,1,\rho}$ with respect to the Borel mass $T$ (a) and the threshold value $\omega_c$ (b) as well as variations of $\bar{\Lambda}_{\Lambda_{c}(2P),1,1,\rho}$ with respect to Borel mass $T$ (c) and the threshold value $\omega_c^\prime$ (d), when $J_{1/2,-,\mathbf{\bar 3_F},1,1,\rho}$ is adopted. In the subfigure~(a) the short-dashed, solid, and long-dashed curves are obtained by fixing $\omega_c=1.10$~GeV, $1.20$~GeV, and $1.30$~GeV, respectively. In the subfigure~(c) the short-dashed, solid, and long-dashed curves are obtained by fixing $\omega_c^\prime=1.47$~GeV, $1.57$~GeV, and $1.67$~GeV, respectively. In the subfigures~(b,d) the short-dashed, solid, and long-dashed curves are gained by setting $T=0.260$~GeV, $0.264$~GeV, and $0.266$~GeV, respectively.}
\label{fig:msc}
\end{center}
\end{figure*}

To provide a clear visualization, we show the variations of CVG, PC$^\prime$, and PC with respect to the Borel mass $T$ in Fig.~\ref{fig:cvgpc}. Additionally, the fluctuations of $\bar{\Lambda}_{\Lambda_c({1\over2}^-,1P)}$ and $\bar{\Lambda}_{\Lambda_c({1\over2}^-,2P)}$ are exhibited with respect to $T$, $\omega_c$, and $\omega_c^\prime$ in Fig.~\ref{fig:msc}. We find that these curves remain sufficiently stable inside the regions $0.260$~GeV$\leq T\leq 0.266$~GeV, $1.10$~GeV$\leq \omega_c \leq 1.30$~GeV, and $1.47$~GeV$\leq \omega_c^\prime \leq 1.67$~GeV, where the associated numerical results are obtained:
\begin{eqnarray}
\bar{\Lambda}_{\Lambda_c({1\over2}^-,1P)}&=&1.22\pm0.06~{\rm GeV}\, ,\\
f_{\Lambda_c({1\over2}^-,1P)}&=&0.043\pm 0.008~{\rm GeV^4}\, ,\\
\bar{\Lambda}_{\Lambda_c({1\over2}^-,2P)}&=&1.40\pm0.07~{\rm GeV}\, ,\\
f_{\Lambda_c({1\over2}^-,2P)}&=&0.057\pm 0.016~{\rm GeV^4}\, .
\end{eqnarray}

\subsection{The sum rules at the $\mathcal{O}(1/m_Q)$ order}

In this subsection we perform analysis at the $\mathcal{O}(1/m_Q)$ order. To do this we utilize the following Lagrangian of HQET:
\begin{eqnarray}
\mathcal{L}=\bar h_v i v\cdot D h_v +{1\over 2m_Q}\mathcal{K}+{1\over 2m_Q}\mathcal{S} \, .
\end{eqnarray}
In this Lagrangian we designate $\mathcal{K}$ to represent the operator of nonrelativistic kinetic energy,
\begin{eqnarray}
\mathcal{K}=\bar h_v (i D_t)^2 h_v\, ,
\label{eq:k}
\end{eqnarray}
and use $\mathcal{S}$ to denote the Pauli term describing the chromomagnetic interaction,
\begin{eqnarray}
\mathcal{S}={g\over2} C_{mag}(m_Q/\mu)\bar h_v \sigma_{\mu\nu} G^{\mu\nu} h_v\, ,
\label{eq:s}
\end{eqnarray}
where $C_{mag}(M_Q/\mu)=[\alpha_s(m_Q)/\alpha_s(\mu)]^{3/\beta_0}$ and $\beta_0=11-2n_f/3$.

\begin{widetext}
The two pole terms arise at the $\mathcal{O}(1/m_Q)$ order when we perform the Taylor expansion for Eq.~(\ref{eq:hadron}), with $\delta m_{\Lambda_c({1\over2}^-,1P)}$, $\delta m_{\Lambda_c({1\over2}^-,2P)}$, $\delta f_{\Lambda_c({1\over2}^-,1P)}$ and $\delta f_{\Lambda_c({1\over2}^-,2P)}$ representing the corrections to $m_{\Lambda_c({1\over2}^-,1P)}$, $m_{\Lambda_c({1\over2}^-,2P)}$, $f_{\Lambda_c({1\over2}^-,1P)}$, and $f_{\Lambda_c({1\over2}^-,2P)}$, respectively:
\begin{eqnarray}
\Pi(\omega,\omega^\prime)_{pole}
 &=&{(f_{\Lambda_c({1\over2}^-,1P)}+\delta f_{\Lambda_c({1\over2}^-,1P)})^2\over \bar \Lambda_{\Lambda_c({1\over2}^-,1P)}+\delta m_{\Lambda_c({1\over2}^-,1P)}-\omega}+{(f_{\Lambda_c({1\over2}^-,2P)}+\delta f_{\Lambda_c({1\over2}^-,2P)})^2\over \bar \Lambda_{\Lambda_c({1\over2}^-,2P)}+\delta m_{\Lambda_c({1\over2}^-,2P)}-\omega^\prime}
\\ \nonumber &=&{f_{\Lambda_c({1\over2}^-,1P)}^2\over \bar \Lambda_{\Lambda_c({1\over2}^-,1P)}-\omega}-{\delta m_{\Lambda_c({1\over2}^-,1P)} f_{\Lambda_c({1\over2}^-,1P)}^2\over (\bar \Lambda_{\Lambda_c({1\over2}^-,1P)}-\omega)^2}+{2f_{\Lambda_c({1\over2}^-,1P)}\delta f_{\Lambda_c({1\over2}^-,1P)}\over \bar \Lambda_{\Lambda_c({1\over2}^-,1P)} -\omega}\\
\nonumber &+&{f_{\Lambda_c({1\over2}^-,2P)}^2\over \bar \Lambda_{\Lambda_c({1\over2}^-,2P)}-\omega^\prime}-{\delta m_{\Lambda_c({1\over2}^-,2P)} f_{\Lambda_c({1\over2}^-,2P)}^2\over (\bar \Lambda_{\Lambda_c({1\over2}^-,2P)}-\omega^\prime)^2}+{2f_{\Lambda_c({1\over2}^-,2P)}\delta f_{\Lambda_c({1\over2}^-,2P)}\over \bar \Lambda_{\Lambda_c({1\over2}^-,2P)} -\omega^\prime} \, ,
\end{eqnarray}
where $\delta m_{\Lambda_c({1\over2}^-,1P)}$ and $\delta m_{\Lambda_c({1\over2}^-,2P)}$ can be calculated through the following three-point correlation functions
\begin{eqnarray}
\delta_O\Pi_{j,P,F,j_l,s_l,\rho/\lambda}^{\alpha_1\cdots \alpha_{j-1/2},\beta_1\cdots \beta_{j-1/2}}
 &=&i^2 \int d^4x d^4 y e^{ik\cdot x-i k^\prime \cdot y}\langle 0|T[J_{i,P,F,j_l,s_l,\rho/\lambda}^{\alpha_1\cdots \alpha_{j-1/2}}(x)O(0)J_{j,P,F,j_l,s_l,\rho/\lambda}^{\beta_1\cdots \beta_{j-1/2}}(y)]|0\rangle
\label{eq:three-point}
\\ \nonumber &=& \mathbb{S}[g_t^{\alpha_1\beta_1}\cdots g_t^{\alpha_{j-1/2}\beta_{j-1/2}}]{1+v\!\!\!\slash\over 2}\delta_O\Pi_{j,P,F,j_l,s_l,\rho/\lambda} \, ,
\end{eqnarray}
with $O=\mathcal{K}$ or $\mathcal{S}$.

We can write Eq.~(\ref{eq:three-point}) at the hadron level as
\begin{eqnarray}
\delta_{\mathcal{K}}\Pi&=&{f_{\Lambda_c({1\over2}^-,1P)}^2 K_{\Lambda_c({1\over2}^-,1P)}\over (\bar \Lambda_{\Lambda_c({1\over2}^-,1P)} -\omega_1)(\bar \Lambda_{\Lambda_c({1\over2}^-,1P)} -\omega_1^\prime)}+{f_{\Lambda_c({1\over2}^-,2P)}^2 K_{\Lambda_c({1\over2}^-,2P)}\over(\bar\Lambda_{\Lambda_c({1\over2}^-,2P)}-\omega_2)(\bar\Lambda_{\Lambda_c({1\over2}^-,2P)}-\omega_2^\prime)}\\
\nonumber &+&{f_{\Lambda_c({1\over2}^-,1P)}f_{\Lambda_c({1\over2}^-,2P)}K^\prime_{\Lambda_c({1\over2}^-)}\over(\bar \Lambda_{\Lambda_c({1\over2}^-,1P)}
-\omega_1)(\bar \Lambda_{\Lambda_c({1\over2}^-,2P)}-\omega_2^\prime)}
+{f_{\Lambda_c({1\over2}^-,2P)}f_{\Lambda_c({1\over2}^-,1P)}K_{\Lambda_c({1\over2}^-)}^{\prime}\over(\bar \Lambda_{\Lambda_c({1\over2}^-,2P)}-\omega_2)(\bar \Lambda_{\Lambda_c({1\over2}^-,1P)}-\omega_1^\prime)}+\cdots \, ,
\\  \delta_{\mathcal{S}}\Pi&=&{d_M f_{\Lambda_c({1\over2}^-,1P)}^2\Sigma_{\Lambda_c({1\over2}^-,1P)}\over (\bar \Lambda_{\Lambda_c({1\over2}^-,1P)}-\omega_1)(\bar\Lambda_{\Lambda_c({1\over2}^-,1P)}-\omega_1^\prime)}+{d_Mf_{\Lambda_c({1\over2}^-,2P)}^2 \Sigma_{\Lambda_c({1\over2}^-,2P)}\over(\bar\Lambda_{\Lambda_c({1\over2}^-,2P)}-\omega_2)(\bar\Lambda_{\Lambda_c({1\over2}^-,2P)}
-\omega_2^\prime)}\\
\nonumber &+&{d_Mf_{\Lambda_c({1\over2}^-,1P)}f_{\Lambda_c({1\over2}^-,2P)}\Sigma_{\Lambda_c({1\over2}^-)}^\prime\over(\bar \Lambda_{\Lambda_c({1\over2}^-,1P)}-\omega_1)(\bar \Lambda_{\Lambda_c({1\over2}^-,2P)}-\omega_2^\prime)}+{d_Mf_{\Lambda_c({1\over2}^-,2P)}f_{\Lambda_c({1\over2}^-,1P)}\Sigma_{\Lambda_c({1\over2}^-)}^{\prime}\over(\bar \Lambda_{\Lambda_c({1\over2}^-,2P)}-\omega_2)(\bar \Lambda_{\Lambda_c({1\over2}^-,1P)}-\omega_1^\prime)}+\cdots \, ,
\end{eqnarray}
where $K_{\Lambda_c({1\over2}^-,1P/2P)}$, $K_{\Lambda_c({1\over2}^-)}^{\prime}$, $\Sigma_{\Lambda_c({1\over2}^-,1P/2P)}$, and $\Sigma_{\Lambda_c({1\over2}^-)}^{\prime}$ are the matrix elements formed from the operators $\mathcal{K}$ and $\mathcal{S}$ sandwiched between the $\Lambda_c({1\over2}^-,1P/2P)$ states:
\begin{eqnarray}
\left(\begin{array}{ll}
\langle \Lambda_c({1\over2}^-,1P)|\mathcal{K}|\Lambda_c({1\over2}^-,1P)\rangle & \langle \Lambda_c({1\over2}^-,1P)|\mathcal{K}|\Lambda_c({1\over2}^-,2P)\\
\langle \Lambda_c({1\over2}^-,2P)|\mathcal{K}|\Lambda_c({1\over2}^-,1P)\rangle & \langle \Lambda_c({1\over2}^-,2P)|\mathcal{K}|\Lambda_c({1\over2}^-,2P)
\end{array}\right)&=& \left(\begin{array}{ll}
K_{\Lambda_c({1\over2}^-,1P)}& K^\prime_{\Lambda_c({1\over2}^-)}\\
K^\prime_{\Lambda_c({1\over2}^-)}& K_{\Lambda_c({1\over2}^-,2P)}
\end{array}\right)\, ,\\
\left(\begin{array}{ll}
\langle \Lambda_c({1\over2}^-,1P)|\mathcal{S}|\Lambda_c({1\over2}^-,1P)\rangle & \langle \Lambda_c({1\over2}^-,1P)|\mathcal{S}|\Lambda_c({1\over2}^-,2P)\\
\langle \Lambda_c({1\over2}^-,2P)|\mathcal{S}|\Lambda_c({1\over2}^-,1P)\rangle &\langle \Lambda_c({1\over2}^-,2P)|\mathcal{S}|\Lambda_c({1\over2}^-,2P)
\end{array}\right)&=& C_{mag}\left(\begin{array}{ll}
d_M \Sigma_{\Lambda_c({1\over2}^-,1P)}& d_M \Sigma^\prime_{\Lambda_c({1\over2}^-)}\\
d_M \Sigma^\prime_{\Lambda_c({1\over2}^-)}& d_M \Sigma_{\Lambda_c({1\over2}^-,2P)}
\end{array}\right)\, ,
\end{eqnarray}
where $d_M=d_{j,j_l}$, $d_{j_l-1/2,j_l}=2j_l+2$, and $d_{j_l+1/2,j_l}=-2j_l$.
\end{widetext}

During the calculations we find that the non-diagonal elements are not zero, but significantly smaller than the diagonal elements. Hence, the contribution of non-diagonal elements is negligible. We note that the term $\mathcal{S}$ can induce a mass splitting within the same doublet, and thus, one can obtain the masses of $\Lambda_c({1\over2}^-/{3\over2}^-,1P/2P)$ through the following expression:
\begin{eqnarray}
\delta m &=&-{1\over 4m_c}(K+d_{j,j_l}C_{mag}\Sigma)\, ,\\
 m&=& m_c + \bar{\Lambda} + \delta m\, ,
\end{eqnarray}
where $m_c$ is the charmed quark mass in the $\rm \overline{MS}$ scheme, $\bar{\Lambda}$ is the sum rule result at the leading order, and $\delta m$ is the sum rule result at the $\mathcal{O}(1/m_Q)$ order.

We can also calculate Eq.~(\ref{eq:three-point}) at the quark-gluon level using the method of operator product expansion (OPE). After inserting Eqs.~(\ref{eq:1/2},\ref{eq:k},\ref{eq:s}) into Eq.~(\ref{eq:three-point}) and making a double Borel transformation for $\omega_1$, $\omega_1^\prime$, $\omega_2$, and $\omega_2^\prime$, we obtain the four Borel parameters $T_1$, $T_2$, $T_3$, and $T_4$. We choose $T_1=T_2=T_3=T_4=2T$, and obtain the following sum rules for $K_{\Lambda_c({1\over2}^-,1P/2P)}$ and $\Sigma_{\Lambda_c({1\over2}^-,1P/2P)}$:
\begin{widetext}
\begin{eqnarray}
&&f_{\Lambda_c({1\over2}^-,1P)}^2 K_{\Lambda_c({1\over2}^-,1P)} e^{-\bar\Lambda_{\Lambda_c({1\over2}^-,1P)}/T}+f_{\Lambda_c({1\over2}^-,2P)}^2 K_{\Lambda_c({1\over2}^-,2P)}e^{-\bar\Lambda_{\Lambda_c({1\over2}^-,2P)}/T}\\
\nonumber &+& 2f_{\Lambda_c({1\over2}^-,1P)} f_{\Lambda_c({1\over2}^-,2P)}K_{\Lambda_c({1\over2}^-)}^\prime e^{-{\bar\Lambda_{\Lambda_c({1\over2}^-,1P)}+\bar\Lambda_{\Lambda_c({1\over2}^-,2P)}\over 2T}}
\\ \nonumber &=&\int_0^{\omega_c^\prime}[-{8\omega^9\over 105\pi^4}+{9\langle g_s^2 G G\rangle \omega^5\over 240\pi^4}]e^{-\omega/T}d\omega -{\langle g_s\bar q  \sigma G q\rangle\langle g_s\bar q \sigma G q\rangle\over16}-{\langle\bar q q\rangle\langle g_s\bar q  \sigma G q\rangle\langle g_s^2 GG\rangle\over 256 T^2}\\
\nonumber && -{\langle g_s\bar q \sigma G q\rangle\langle g_s\bar q \sigma G q\rangle\langle g_s^2 GG\rangle\over 4096 T^4}
\\ \nonumber &=&\int_0^{\omega_c}[-{8\omega^9\over 105\pi^4}+{9\langle g_s^2 G G\rangle \omega^5\over 240\pi^4}]e^{-\omega/T}d\omega -{\langle g_s\bar q  \sigma G q\rangle\langle g_s\bar q \sigma G q\rangle\over16}-{\langle\bar q q\rangle\langle g_s\bar q \sigma G q\rangle\langle g_s^2 GG\rangle\over 256 T^2}\\ \nonumber &&-{\langle g_s\bar q \sigma G q\rangle\langle g_s\bar q \sigma G q\rangle\langle g_s^2 GG\rangle\over 4096 T^4}
+  \int_{\omega_c}^{\omega_c^\prime}[-{8\omega^9\over 105\pi^4}+{9\langle g_s^2 G G\rangle \omega^5\over 240\pi^4}]e^{-\omega/T}d\omega
\\ \nonumber &=& \delta_{\mathcal{K}}\Pi_{\Lambda_c({1\over2}^-,1P)}(\omega_c,T)+\delta_{\mathcal{K}}\Pi_{\Lambda_c({1\over2}^-,2P)}(\omega_c,\omega_c^\prime,T)\, ,
\\
&&f_{\Lambda_c({1\over2}^-,1P)}^2 d_M \Sigma_{\Lambda_c({1\over2}^-,1P)} e^{-\bar\Lambda_{\Lambda_c({1\over2}^-,1P)}/T}+f_{\Lambda_c({1\over2}^-,2P)}^2 d_M\Sigma_{\Lambda_c({1\over2}^-,2P)}e^{-\bar\Lambda_{\Lambda_c({1\over2}^-,2P)}/T}\\
\nonumber &+&2f_{\Lambda_c({1\over2}^-,1P)} f_{\Lambda_c({1\over2}^-,2P)}d_M\Sigma_{\Lambda_c({1\over2}^-)}^\prime e^{-{\bar\Lambda_{\Lambda_c({1\over2}^-,1P)}+\bar\Lambda_{\Lambda_c({1\over2}^-,2P)}\over 2T}}
\\ \nonumber &=&\int_0^{\omega_c^\prime}[{3\langle g_s^2 G G\rangle \omega^5\over 120\pi^4}]e^{-\omega/T}d\omega
\\ \nonumber &=&\int_0^{\omega_c}[{3\langle g_s^2 G G\rangle \omega^5\over 120\pi^4}]e^{-\omega/T}d\omega
+ \int_{\omega_c}^{\omega_c^\prime}[{3\langle g_s^2 G G\rangle \omega^5\over 120\pi^4}]e^{-\omega/T}d\omega
\\ \nonumber &=& \delta_{\mathcal{S}}\Pi_{\Lambda_c({1\over2}^-,1P)}(\omega_c,T)+\delta_{\mathcal{S}}\Pi_{\Lambda_c({1\over2}^-,2P)}(\omega_c,\omega_c^\prime,T)\, ,
\end{eqnarray}
where $K_{\Lambda_c({1\over2}^-,1P)}$, $K_{\Lambda_c({1\over2}^-,2P)}$, $S_{\Lambda_c({1\over2}^-,1P)}$, and $S_{\Lambda_c({1\over2}^-,2P)}$ can be extracted as:
\begin{eqnarray}
K_{\Lambda_c({1\over2}^-,1P)}&=&{\delta_{\mathcal{K}}\Pi_{\Lambda_c({1\over2}^-,1P)}\over \Pi_{\Lambda_c({1\over2}^-,1P)}}\, ,~~~~~~d_M\Sigma_{\Lambda_c({1\over2}^-,1P)} ~=~ {\delta_{\mathcal{S}}\Pi_{\Lambda_c({1\over2}^-,1P)}\over \Pi_{\Lambda_c({1\over2}^-,1P)}}\, ,
\\
K_{\Lambda_c({1\over2}^-,2P)}&=&{2 T^2{\partial \delta_{\mathcal{K}}\Pi_{\Lambda_c({1\over2}^-,2P)}\over\partial T}-(\bar \Lambda_{\Lambda_c({1\over2}^-,1P)}+\bar \Lambda_{\Lambda_c({1\over2}^-,2P)})\delta_{\mathcal{K}}\Pi_{\Lambda_c({1\over2}^-,2P)}\over \Pi_{\Lambda_c({1\over2}^-,2P)}(\bar \Lambda_{\Lambda_c({1\over2}^-,2P)}-\bar \Lambda_{\Lambda_c({1\over2}^-,1P)})}\, ,
\\
d_M\Sigma_{\Lambda_c({1\over2}^-,2P)}&=&{2 T^2{\partial \delta_{\mathcal{S}}\Pi_{\Lambda_c({1\over2}^-,2P)}\over\partial T}-(\bar \Lambda_{\Lambda_c({1\over2}^-,1P)}+\bar \Lambda_{\Lambda_c({1\over2}^-,2P)})\delta_{\mathcal{S}}\Pi_{\Lambda_c({1\over2}^-,2P)}\over \Pi_{\Lambda_c({1\over2}^-,2P)}(\bar \Lambda_{\Lambda_c({1\over2}^-,2P)}-\bar \Lambda_{\Lambda_c({1\over2}^-,1P)})}\, .
\end{eqnarray}
\end{widetext}

\begin{figure*}[hbt]
\begin{center}
\subfigure[]{
\scalebox{0.5}{\includegraphics{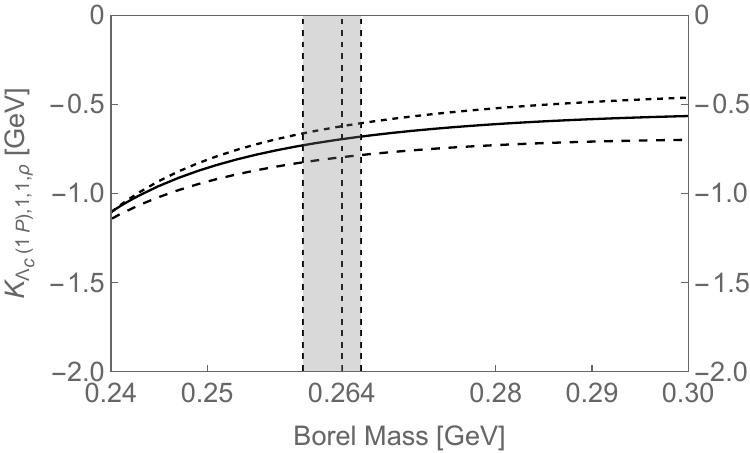}}}~~~~
\subfigure[]{
\scalebox{0.5}{\includegraphics{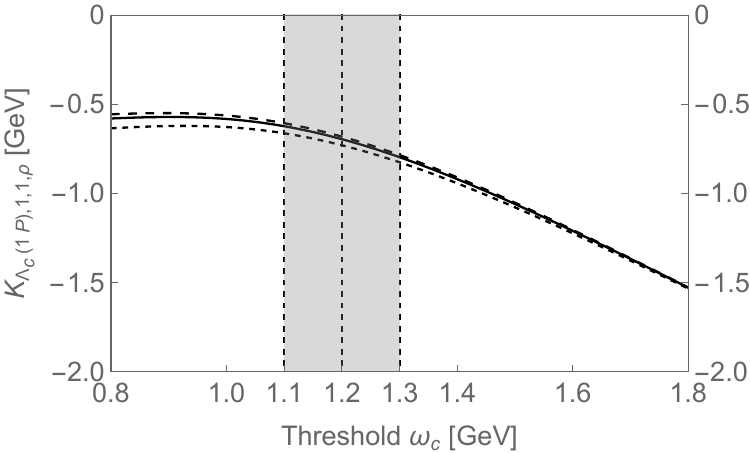}}}~~~~
\\
\subfigure[]{
\scalebox{0.5}{\includegraphics{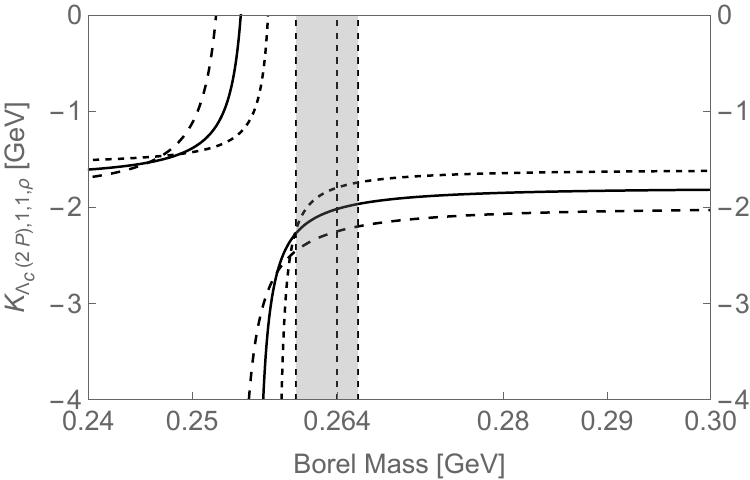}}}~~~~
\subfigure[]{
\scalebox{0.5}{\includegraphics{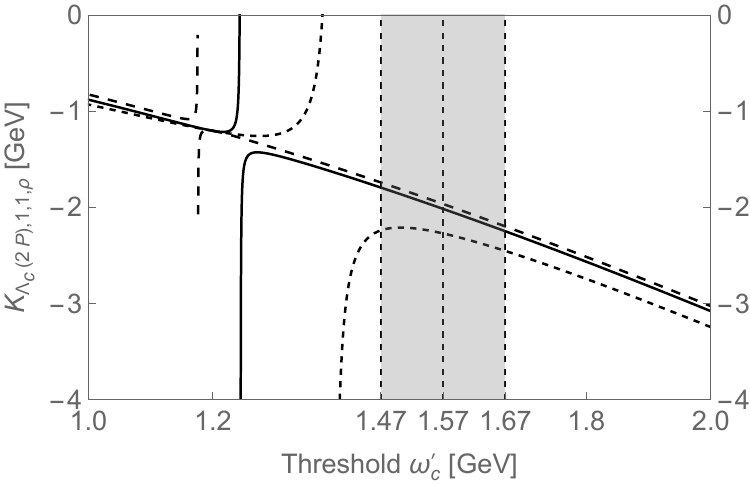}}}~~~~
\caption{Variations of $K_{\Lambda_{c}(1P),1,1,\rho}$ with respect to the Borel mass $T$ (a) and the threshold value $\omega_c$ (b) as well as variations of $K_{\Lambda_{c}(2P),1,1,\rho}$ with respect to the Borel mass $T$ (c) and the threshold value $\omega_c^\prime$ (d), when $J_{1/2,-,\mathbf{\bar 3_F},1,1,\rho}$ is adopted. In the subfigure~(a) the short-dashed, solid, and long-dashed curves are obtained by fixing $\omega_c=1.10$~GeV, $1.20$~GeV, and $1.30$~GeV, respectively. In the subfigure~(c) the short-dashed, solid, and long-dashed curves are obtained by fixing $\omega_c^\prime=1.47$~GeV, $1.57$~GeV, and $1.67$~GeV, respectively. In the subfigures (b,d) the short-dashed, solid, and long-dashed curves are gained by setting $T=0.260$~GeV, $0.264$~GeV, and $0.266$~GeV, respectively.}
\label{fig:k}
\end{center}
\end{figure*}

\begin{figure*}[hbt]
\begin{center}
\subfigure[]{
\scalebox{0.5}{\includegraphics{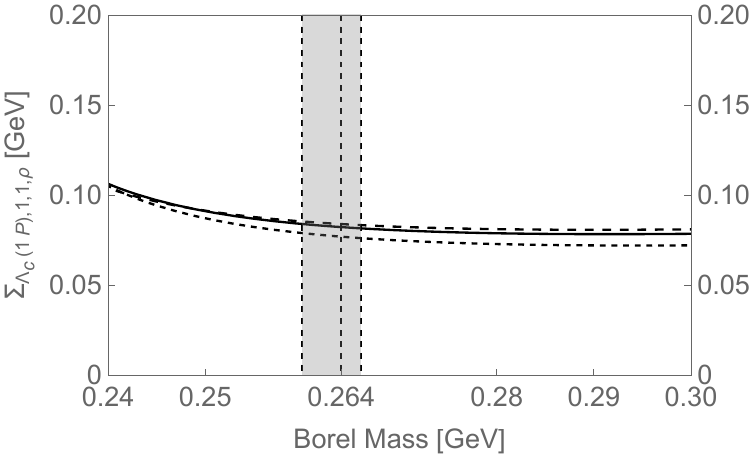}}}~~~~
\subfigure[]{
\scalebox{0.5}{\includegraphics{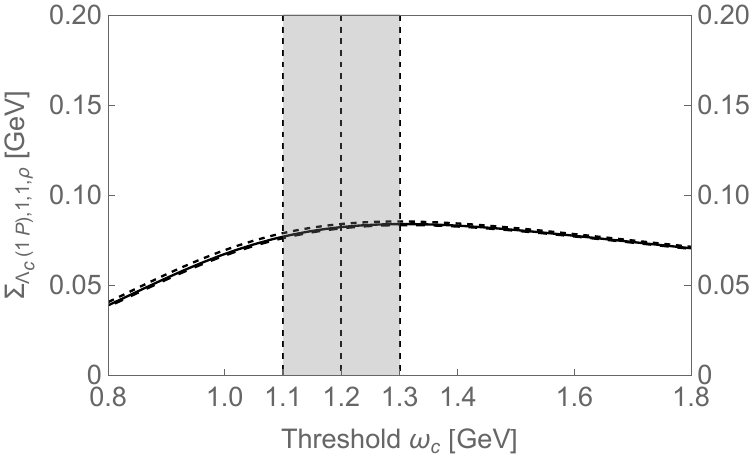}}}~~~~
\\
\subfigure[]{
\scalebox{0.5}{\includegraphics{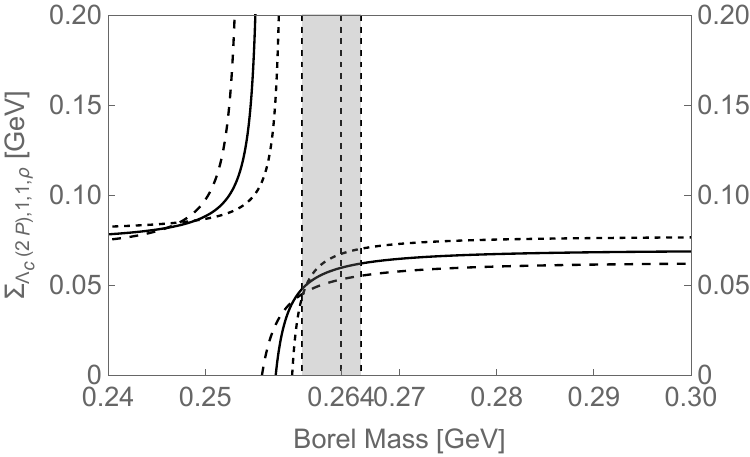}}}~~~~
\subfigure[]{
\scalebox{0.5}{\includegraphics{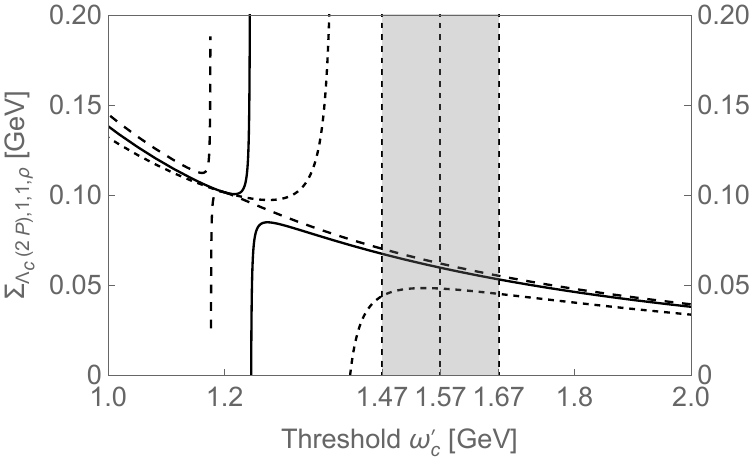}}}~~~~
\caption{Variations of $\Sigma_{\Lambda_{c}(1P),1,1,\rho}$ with respect to the Borel mass $T$ (a) and the threshold value $\omega_c$ (b) as well as variations of $\Sigma_{\Lambda_{c}(1P),1,1,\rho}$ with respect to the Borel mass $T$ (c) and the threshold value $\omega_c^\prime$ (d), when $J_{1/2,-,\mathbf{\bar 3_F},1,1,\rho}$ is adopted. In the subfigure~(a) the short-dashed, solid, and long-dashed curves are obtained by fixing $\omega_c=1.10$~GeV, $1.20$~GeV, and $1.30$~GeV, respectively. In the subfigure~(c) the short-dashed, solid, and long-dashed curves are obtained by fixing $\omega_c^\prime=1.47$~GeV, $1.57$~GeV, and $1.67$~GeV, respectively. In the subfigures~(b,d) the short-dashed, solid, and long-dashed curves are gained by setting $T=0.260$~GeV, $0.264$~GeV, and $0.266$~GeV, respectively.}
\label{fig:s}
\end{center}
\end{figure*}

Their variations are depicted in Fig.~\ref{fig:k} and Fig.~\ref{fig:s} with respect to the Borel mass $T$ as well as the threshold values $\omega_c$ and $\omega_c^\prime$. We find that their dependence on $T$, $\omega_c$, and $\omega_c^\prime$ remains weak in the working regions $0.260$~GeV$\leq T\leq 0.266$~GeV, $1.10$~GeV$\leq \omega_c \leq 1.30$~GeV, and $1.47$~GeV$\leq \omega_c^\prime \leq 1.67$~GeV, where the corresponding numerical results are obtained:
\begin{eqnarray}
K_{\Lambda_c({1\over2}^-,1P)}= -0.69\pm0.10 {\rm GeV}\, ,\\
d_M \Sigma_{\Lambda_c({1\over2}^-,1P)}=0.08\pm0.03{\rm GeV}\, ,
\\ K_{\Lambda_c({1\over2}^-,2P)}=-2.01\pm 0.35{\rm GeV}\, ,\\
d_M\Sigma_{\Lambda_c({1\over2}^-,2P)}=0.06\pm 0.02{\rm GeV}\, .
\end{eqnarray}

\begin{figure*}[hbt]
\begin{center}
\subfigure[]{
\scalebox{0.45}{\includegraphics{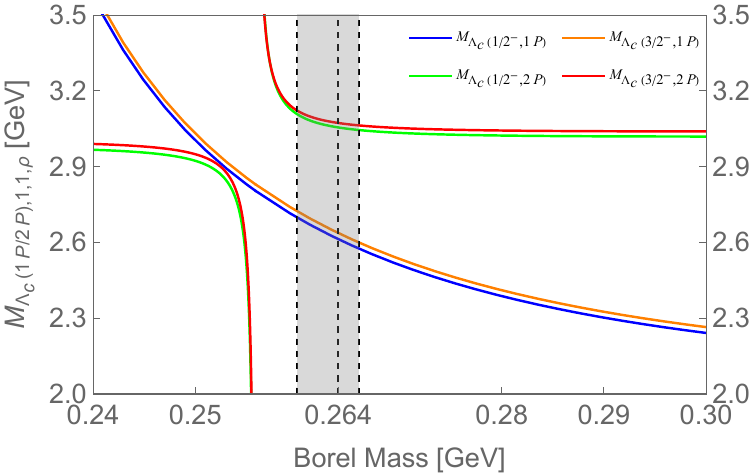}}}~~~~
\subfigure[]{
\scalebox{0.45}{\includegraphics{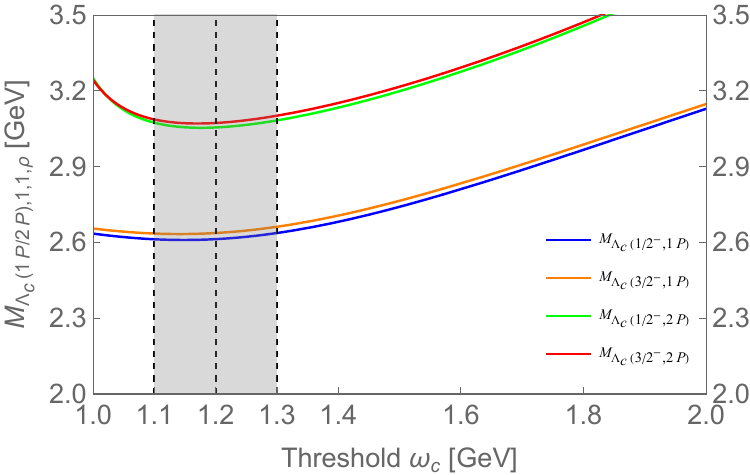}}}~~~~
\subfigure[]{
\scalebox{0.45}{\includegraphics{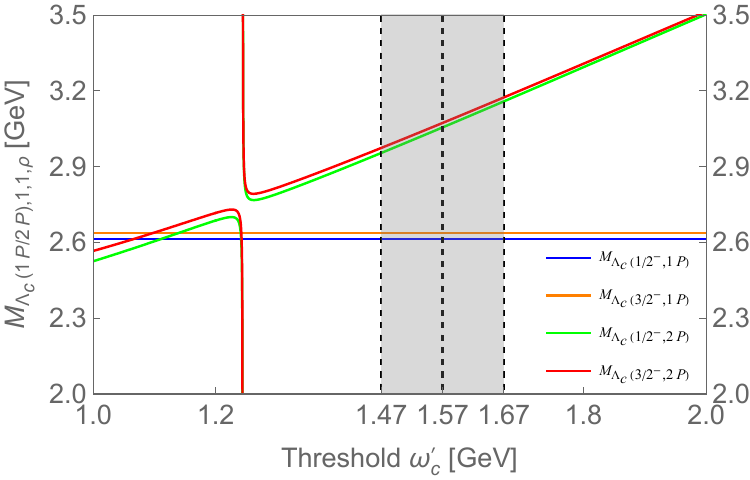}}}
\caption{Variations of $m_{\Lambda_{c}(1P),1,1,\rho}$ and $m_{\Lambda_{c}(2P),1,1,\rho}$ with respect to Borel mass $T$ (a) as well as the threshold values $\omega_c$ (b) and $\omega_c^\prime$ (c), calculated using the charmed baryon doublet $[\Lambda_c,\mathbf{\bar 3_F},1,1,\rho]$. The blue, orange, green, and red curves represent the mass functions of $\Lambda_c({1\over2}^-,1P)$, $\Lambda_c({3\over2}^-,1P)$, $\Lambda_c({1\over2}^-,2P)$ and $\Lambda_c({3\over2}^-,2P)$, respectively. The curves in the left panel are obtained by fixing $\omega_c=1.20$~GeV and $\omega_c^\prime=1.57$~GeV, respectively. The curves in the middle panel are obtained by fixing $T=0.264$~GeV and $\omega_c^\prime=1.47$~GeV, respectively. The curves in the right panel are gained by setting $T=0.264$~GeV and $\omega_c=1.20$~GeV, respectively.}
\label{fig:m}
\end{center}
\end{figure*}

Finally, we can obtain the masses of $\Lambda_c({1\over2}^-/{3\over2}^-,1P/2P)$ belonging to the $[\mathbf{\bar 3}_F, 1, 1, \rho]$ doublet:
\begin{eqnarray}
m_{\Lambda_c({1\over2}^-,1P)}&=& 2.61\pm 0.07 {\rm GeV}\, ,\\
m_{\Lambda_c({3\over2}^-,1P)}&=& 2.63\pm 0.07 {\rm GeV}\, ,\\
m_{\Lambda_c({3\over2}^-,1P)}-m_{\Lambda_c({1\over2}^-,1P)}&=&24\pm 9{\rm MeV}\, ,\\
m_{\Lambda_c({1\over2}^-,2P)}&=& 3.05\pm 0.12 {\rm GeV}\, ,\\
m_{\Lambda_c({3\over2}^-,2P)}&=& 3.07\pm 0.11{\rm GeV}\, ,\\
m_{\Lambda_c({3\over2}^-,2P)}-m_{\Lambda_c({1\over2}^-,2P)}&=&18\pm 7{\rm MeV}\, .
\end{eqnarray}
In Fig.~\ref{fig:m} we present the variations of $m_{\Lambda_{c}(1P),1,1,\rho}$ and $m_{\Lambda_{c}(2P),1,1,\rho}$ with respect to $T$, $\omega_c$, and $\omega_c^\prime$. It is noteworthy that there exist significant theoretical uncertainties in our results for the masses of the charmed baryons, but their differences within the same doublet are determined with less uncertainty since they do not depend much on the charm quark mass.

The results obtained are summarized in Table~\ref{tab:pwavebparameter}. We utilize the charmed baryons belonging to the $[\mathbf{\bar 3}_F, 1, 1, \rho]$ doublet to fit the $\Lambda_c(2595)(J^P={1\over2}^-,1P)$, $\Lambda_c(2625)(J^P={3\over2}^-,1P)$, $\Lambda_c(2910)(J^P={1\over2}^-,2P)$, $\Lambda_c(2940)(J^P={3\over2}^-,2P)$, $\Xi_c(2790)(J^P={1\over2}^-,1P)$, $\Xi_c(2815)(J^P={3\over2}^-,1P)$, and $\Xi_c(3123)(J^P={3\over2}^-,2P)$. Within our QCD sum rule framework, the mass spectra of $1P$-wave charmed baryons are consistent with the experimental measurements, while the mass spectra of $2P$-wave charmed baryons are higher than experimental measurements, which align with the results in quark models. Apart from the mass spectra, we also need to analyze their decay properties. The decay constant $f$ is an important input parameter, which will be used to study the decay widths of $\Lambda_c({1\over2}^-/{3\over2}^-,1P/2P)$ and $\Xi_c({1\over2}^-/{3\over2}^-,1P/2P)$ in the next section. For completeness, we have also investigated the $1P$ and $2P$-wave charmed baryons belonging to the $[\mathbf{\bar 3}_F,1,0,\lambda]$ doublet using the following interpolating currents:
\begin{eqnarray}
&&J_{1/2,-,\mathbf{\bar 3_F},1,0,\lambda}\\
\nonumber&=&i\epsilon_{abc}\big([\mathcal{D}_t^\mu q^{aT}]\mathcal{C}\gamma_5 q^b+q^{aT}\mathcal{C}\gamma_5[\mathcal{D}_t^\mu q^b]\big)\gamma_t^\mu\gamma_5h_v^c,\\
&&J_{3/2,-,\mathbf{\bar 3_F},1,0,\lambda}^\alpha\\
\nonumber &=&i\epsilon_{abc}\big([\mathcal{D}_t^\mu q^{aT}]\mathcal{C}\gamma_5 q^b+q^{aT}\mathcal{C}\gamma_5[\mathcal{D}_t^\mu q^b]\big)\big(g_t^{\alpha\mu}-{1\over3}\gamma_t^\alpha\gamma_t^\mu\big)h_v^c \, .
\end{eqnarray}

The obtained results are summarized in Appendix~\ref{sec:charmedsumrule}. We find that the $\rho$-mode doublet $[\mathbf{\bar 3}_F, 1, 1, \rho]$ appears to be lower than the $\lambda$-mode doublet $[\mathbf{\bar 3}_F,1,0,\lambda]$. This behavior aligns with our previous QCD sum rule results for their corresponding doublets of the $SU(3)$ flavor $\mathbf{6}_F$~\cite{Yang:2021lce,Yang:2020zrh,Yang:2020zjl}, but contradicts the expectation from the quark model~\cite{Yoshida:2015tia,Nagahiro:2016nsx}. However, this discrepancy could be attributed to the significant uncertainties associated with the mass difference between different multiplets within our QCD sum rule framework. 

\begin{table*}[hbtp]
\begin{center}
\renewcommand{\arraystretch}{1.5}
\caption{Parameters of the $1P$- and $2P$-wave charmed baryons belonging to the $[\mathbf{\bar 3}_F, 1, 1, \rho]$ doublets, calculated using the method of QCD sum rules within the framework of heavy quark effective theory. Decay constants in the last column satisfy $f_{\Xi_c^+} = f_{\Xi_c^0}$.}
\begin{tabular}{ c | c | c | c | c c | c | c}
\hline\hline
\multirow{2}{*}{~~B~~} & $\omega_c$ & ~~~Working region~~~ & ~~~~~~~$\overline{\Lambda}$~~~~~~~ & ~~~Baryon~~~ & ~~~~Mass~~~~~ & ~Difference~ & Decay constant
\\                                               & (GeV) & (GeV)      & (GeV)                &                               ($j^P$)       & (GeV)      & (MeV)        & (GeV$^{4}$)
\\ \hline\hline
\multirow{2}{*}{$\Lambda_c(1P)$} & \multirow{2}{*}{1.20} & \multirow{2}{*}{$0.260\le T \le 0.266$} & \multirow{2}{*}{$1.22 \pm 0.06$} & $\Lambda_c(1/2^-)$ & $2.61\pm 0.07$ & \multirow{2}{*}{$24 \pm 9$} & $0.043 \pm 0.008~(\Lambda^+_c(1/2^-))$
\\ \cline{5-6}\cline{8-8}
  & & & & $\Lambda_c(3/2^-)$ & $
2.63\pm 0.07$ & &$0.020 \pm 0.004~(\Lambda^+_c(3/2^-))$
\\ \cline{1-8}
 \multirow{2}{*}{$\Xi_c(1P)$} & \multirow{2}{*}{1.45} & \multirow{2}{*}{$0.248\le T \le 0.257$} & \multirow{2}{*}{$1.43 \pm 0.13$} & $\Xi_c(1/2^-)$ & $2.78\pm 0.14$ & \multirow{2}{*}{$29 \pm 10$} & $0.069 \pm 0.023~(\Xi_c^0(1/2^-))$
\\ \cline{5-6}\cline{8-8}
  & & & & $\Xi_c(3/2^-)$ & $2.81\pm 0.14$ & &$0.033 \pm 0.011~(\Xi_c^0(3/2^-))$
\\ \cline{1-8}
\multirow{2}{*}{$\Lambda_c(2P)$} & \multirow{2}{*}{1.57} & \multirow{2}{*}{$0.260\leq T \leq 0.266$} & \multirow{2}{*}{$1.40 \pm 0.07$} & $\Lambda_c(1/2^-)$ & $3.05\pm 0.12$ & \multirow{2}{*}{$18 \pm 7$} & $0.057 \pm 0.016~(\Lambda^+_c(1/2^-))$
\\ \cline{5-6}\cline{8-8}
  & & & & $\Lambda_c(3/2^-)$ & $3.07\pm 0.11$ & &$0.027 \pm 0.007~(\Lambda^+_c(3/2^-))$
\\ \cline{1-8}
 \multirow{2}{*}{$\Xi_c(2P)$} & \multirow{2}{*}{1.60} & \multirow{2}{*}{$0.248\le T \le 0.257$} & \multirow{2}{*}{$1.53 \pm 0.07$} & $\Xi_c(1/2^-)$ & $3.16\pm 0.12$ & \multirow{2}{*}{$18 \pm 7$} & $0.048 \pm 0.026~(\Xi_c^+(1/2^-))$
\\ \cline{5-6}\cline{8-8}
 & & & & $\Xi_c(3/2^-)$ & $3.18\pm 0.12$ & &$0.022 \pm 0.012~(\Xi_c^+(3/2^-))$
\\ \hline\hline
\end{tabular}
\label{tab:pwavebparameter}
\end{center}
\end{table*}


\section{Decay analyses from light-cone sum rules}
\label{sec:decay}

In this section we investigate the decay properties of $\Lambda_c(1P/2P)$ and $\Xi_c(1P/2P)$ belonging to the $[\mathbf{\bar 3}_F, 1, 1, \rho]$ doublet. To achieve this we apply the light-cone sum rule method to study their $S$- and $D$-wave decays into the ground-state charmed baryons and light pseudoscalar/vector mesons. The relevant decay channels include:
\begin{widetext}
\begin{eqnarray}
&(a1)& {\bf \Gamma\Big[} \Lambda_c[1/2^-,1P/2P] \rightarrow \Lambda_c + \pi {\Big]}
=  {\bf \Gamma\Big[}\Lambda_c^+[1/2^-,1P/2P] \rightarrow \Lambda_c^+ +\pi^0 {\Big]} \, ,
\\ &(a2)& {\bf \Gamma\Big[} \Lambda_c[1/2^-,1P] \rightarrow \Sigma_c + \pi {\Big]}
={\bf \Gamma\Big[} \Lambda_c^+[1/2^-,1P] \rightarrow \Sigma_c^{+}+\pi^0 {\Big]}\\
\nonumber &&~~~~~~~~~~~~~~~~~~~~~~~~~~~~~~~~~~~+2\times {\bf \Gamma\Big[} \Lambda_c^+[1/2^-,1P] \rightarrow \Sigma_c^{++}+\pi^-\to\Lambda_c^++\pi^++\pi^- {\Big]} \, ,
\\ &(a3)& {\bf \Gamma\Big[}\Lambda_c[1/2-,2P]\rightarrow \Sigma_c +\pi{\Big]}
=3\times {\bf\Gamma\Big[}\Lambda_c[1/2-,2P]\rightarrow \Sigma_c^{++} +\pi^-{\Big]}\, ,
\\ &(a4)& {\bf \Gamma\Big[} \Lambda_c[1/2^-,1P] \rightarrow \Sigma_c^* + \pi\rightarrow\Lambda_c+\pi+\pi {\Big]}
= 3 \times {\bf \Gamma \Big[}\Lambda_c^+[1/2^-,1P] \rightarrow \Sigma_c^{*++}+\pi^-\rightarrow \Lambda_c^++\pi^++\pi^-{\Big ]} \, ,
\\ &(a5)& {\bf \Gamma\Big[} \Lambda_c[1/2^-,2P] \rightarrow \Sigma_c^* + \pi {\Big]}
= 3 \times {\bf \Gamma \Big[}\Lambda_c^+[1/2^-,2P] \rightarrow \Sigma_c^{*++}+\pi^-\, ,
\\ &(a6)& {\bf\Gamma\Big[} \Lambda_c[1/2^-,1P/2P] \rightarrow \Lambda_c + \rho \rightarrow\Lambda_c+\pi+\pi{\Big ]}
= {\bf \Gamma\Big[} \Lambda_c^+[1/2^-,1P/2P] \rightarrow \Lambda_c^+ +\pi^++ \pi^- {\Big ]} \, ,
\\ &(a7)& { \bf\Gamma\Big[}\Lambda_c[1/2^-,1P/2P] \rightarrow \Sigma_c + \rho\rightarrow\Sigma_c+\pi+\pi{\Big ]}
= 3 \times { \bf\Gamma\Big[}\Lambda_c^+[1/2^-,1P/2P] \rightarrow \Sigma_c^+ +\pi^++ \pi^-{\Big ]} \, ,
\\ &(a8)&{\bf \Gamma\Big[}\Lambda_c[1/2^-,1P/2P] \rightarrow \Sigma_c^* + \rho\rightarrow\Sigma_c^*+\pi+\pi{\Big ]}
= 3 \times { \bf\Gamma\Big[}\Lambda_c^+[1/2^-,1P/2P] \rightarrow \Sigma_c^{*+} +\pi^++ \pi^-{\Big ]} \, ,
\\ &(b1)& {\bf \Gamma\Big[}\Lambda_c[3/2^-,1P/2P] \rightarrow \Lambda_c + \pi{\Big ]}
= {\bf \Gamma\Big[}\Lambda_c^+[3/2^-] \rightarrow \Lambda_c^+ +\pi^0{\Big ]} \, ,
\\ &(b2)&{\bf \Gamma\Big[}\Lambda_c[3/2^-,1P] \rightarrow \Sigma_c + \pi\rightarrow\Lambda_c+\pi+\pi{\Big ]}
= 3 \times {\bf \Gamma\Big[}\Lambda_c^+[3/2^-,1P] \rightarrow \Sigma_c^{++} +\pi^-\rightarrow\Lambda_c^++\pi^++\pi^-{\Big ]} \, ,
\\  &(b3)&{\bf \Gamma\Big[}\Lambda_c[3/2^-,2P] \rightarrow \Sigma_c + \pi{\Big ]}
= 3 \times {\bf \Gamma\Big[}\Lambda_c^+[3/2^-,2P] \rightarrow \Sigma_c^{++} +\pi^-{\Big ]} \, ,
\\ &(b4)&{\bf \Gamma\Big[}\Lambda_c[3/2^-,1P] \rightarrow \Sigma_c^* + \pi\rightarrow\Lambda_c+\pi+\pi{\Big ]}
= 3 \times {\bf \Gamma\Big[}\Lambda_c^+[3/2^-,1P] \rightarrow \Sigma_c^{*++} +\pi^-\rightarrow\Lambda_c^++\pi^+\pi^-{\Big ]} \, ,
\\ &(b5)&{\bf \Gamma\Big[}\Lambda_c[3/2^-,2P] \rightarrow \Sigma_c^* + \pi{\Big ]}
= 3 \times {\bf \Gamma\Big[}\Lambda_c^+[3/2^-,2P] \rightarrow \Sigma_c^{*++} +\pi^-{\Big ]} \, ,
\\ &(b6)&{\bf \Gamma\Big[} \Lambda_c[3/2^-,1P/2P] \rightarrow \Lambda_c + \rho \rightarrow\Lambda_c+\pi+\pi{\Big ]}
= { \bf\Gamma\Big[}\Lambda_c^+[3/2^-,1P/2P] \rightarrow \Lambda_c^+ +\pi^++ \pi^- {\Big ]} \, ,
\\ &(b7)& { \bf\Gamma\Big[}\Lambda_c[3/2^-,1P/2P] \rightarrow \Sigma_c + \rho\rightarrow\Sigma_c+\pi+\pi{\Big ]}
= 3 \times { \bf\Gamma\Big[}\Lambda_c^+[3/2^-,1P/2P] \rightarrow \Sigma_c^+ +\pi^++ \pi^-{\Big ]} \, ,
\\&(b8)& { \bf\Gamma\Big[}\Lambda_c[3/2^-,1P/2P] \rightarrow \Sigma_c^* + \rho\rightarrow\Sigma_c^*+\pi+\pi {\Big ]}
= 3 \times {\bf \Gamma\Big[}\Lambda_c^0[3/2^-,1P/2P] \rightarrow \Sigma_c^{*+} + \pi^++\pi^- {\Big ]} \, ,
\\ &(c1)& {\bf \Gamma\Big[}\Xi_c[1/2^-,1P] \rightarrow \Lambda_c + \bar K{\Big ]}
= {\bf \Gamma\Big[}\Xi_c^0[1/2^-,1P] \rightarrow\Lambda_c^+ +K^-{\Big ]} \, ,
\\ &(c2)&{\bf \Gamma\Big[}\Xi_c[1/2^-,1P] \rightarrow\Xi_c + \pi{\Big ]}
= {3\over2} \times {\bf \Gamma\Big[}\Xi_c^0[1/2^-,1P] \rightarrow \Xi_c^+ +\pi^-{\Big ]} \, ,
\\ &(c3)&{\bf \Gamma\Big[}\Xi_c[1/2^-,1P] \rightarrow\Sigma_c + \bar K{\Big ]}
= 3 \times {\bf \Gamma\Big[}\Xi_c^0[1/2^-,1P] \rightarrow \Sigma_c^+ +K^-{\Big ]} \, ,
\\ &(c4)& { \bf\Gamma\Big[}\Xi_c[1/2^-,1P] \rightarrow \Xi_c^{\prime}+\pi {\Big]}
= {3\over2}\times{\bf \Gamma\Big[}\Xi_c^0[1/2^-,1P]\rightarrow\Xi_c^{\prime+}+\pi^-{\Big]} \, ,
\\ &(c5)& {\bf \Gamma\Big[}\Xi_c[1/2^-,1P] \rightarrow \Sigma_c^* + K{\Big ]}
= 3 \times {\bf \Gamma\Big[}\Xi_c^0[1/2^-,1P] \rightarrow \Sigma_c^{*+} + K^-{\Big ]} \, ,
\\ &(c6)& {\bf \Gamma\Big[}\Xi_c[1/2^-,1P] \rightarrow \Xi_c^* + \pi\rightarrow \Xi_c+\pi+\pi {\Big ]}
= {9 \over 2} \times {\bf \Gamma\Big[}\Xi_c^{0}[1/2^-,1P] \rightarrow\Xi_c^{*+} + \pi^-\rightarrow\Xi_c^++\pi^0+\pi^-{\Big]} \, ,
\\ &(c7)& {\bf \Gamma\Big[}\Xi_c[1/2^-,1P] \rightarrow \Lambda_c + \bar{K}^*\rightarrow\Lambda_c+\bar K+\pi{\Big ]}
=3\times  {\bf \Gamma\Big[}\Xi_c^0[1/2^-,1P] \rightarrow \Lambda_c^+ +  K^- +\pi^0{\Big ]} \, ,
\\ &(c8)& {\bf \Gamma\Big[}\Xi_c[1/2^-,1P] \rightarrow\Xi_c + \rho\rightarrow\Xi_c+\pi+\pi{\Big ]}
= {3\over2} \times {\bf \Gamma\Big[}\Xi_c^0[1/2^-,1P] \rightarrow \Xi_c^+ + \pi^0+\pi^-{\Big ]} \, ,
\\ &(c9)& {\bf \Gamma\Big[}\Xi_c[1/2^-,1P] \rightarrow \Sigma_c^* + \bar K^*\rightarrow\Sigma_c^{*}+\bar K+\pi{\Big ]}
= 9 \times {\bf \Gamma\Big[}\Xi_c^0[1/2^-,1P] \rightarrow \Sigma_c^{*+} + K^-+ \pi^0{\Big ]} \, ,
\\ &(c10)& {\bf \Gamma\Big[}\Xi_c[1/2^-,1P] \rightarrow \Xi_c^{*}+ \rho\rightarrow\Xi_c^{*}+\pi+\pi {\Big ]}
={3\over2} \times {\bf \Gamma\Big[}\Xi_c^0[1/2^-,1P] \rightarrow \Xi_c^{*+} + \pi^0+\pi^- {\Big ]}\, ,
\\ &(c11)& {\bf \Gamma\Big[}\Xi_c[1/2^-,2P] \rightarrow \Lambda_c + \bar K{\Big ]}
= {\bf \Gamma\Big[}\Xi_c^+[1/2^-,2P] \rightarrow\Lambda_c^+ +\bar{K}^0{\Big ]} \, ,
\\ &(c12)&{\bf \Gamma\Big[}\Xi_c[1/2^-,2P] \rightarrow\Xi_c + \pi{\Big ]}
= {3\over2} \times {\bf \Gamma\Big[}\Xi_c^+[1/2^-,2P] \rightarrow \Xi_c^0 +\pi^+{\Big ]} \, ,
\\ &(c13)&{\bf \Gamma\Big[}\Xi_c[1/2^-,2P] \rightarrow\Sigma_c + \bar K{\Big ]}
= {3\over2} \times {\bf \Gamma\Big[}\Xi_c^+[1/2^-,2P] \rightarrow \Sigma_c^{++} +K^-{\Big ]} \, ,
\\ &(c14)& { \bf\Gamma\Big[}\Xi_c[1/2^-,2P] \rightarrow \Xi_c^{\prime}+\pi {\Big]}
= {3\over2}\times{\bf \Gamma\Big[}\Xi_c^+[1/2^-,2P]\rightarrow\Xi_c^{\prime0}+\pi^+{\Big]} \, ,
\\ &(c15)& {\bf \Gamma\Big[}\Xi_c[1/2^-,2P] \rightarrow \Sigma_c^* + K{\Big ]}
= {3\over2} \times {\bf \Gamma\Big[}\Xi_c^+[1/2^-,2P] \rightarrow \Sigma_c^{*++} + K^-{\Big ]} \, ,
\\ &(c16)& {\bf \Gamma\Big[}\Xi_c[1/2^-,2P] \rightarrow \Xi_c^* + \pi{\Big ]}
= {3\over2} \times {\bf \Gamma\Big[}\Xi_c^{+}[1/2^-,2P] \rightarrow\Xi_c^{*0} + \pi^+{\Big]} \, ,
\\ &(c17)& {\bf \Gamma\Big[}\Xi_c[1/2^-,2P] \rightarrow \Lambda_c + \bar{K}^*\rightarrow\Lambda_c+\bar K+\pi{\Big ]}
=3\times  {\bf \Gamma\Big[}\Xi_c^+[1/2^-,2P] \rightarrow \Lambda_c^+ +  \bar{K}^0 +\pi^0{\Big ]} \, ,
\\ &(c18)& {\bf \Gamma\Big[}\Xi_c[1/2^-,2P] \rightarrow\Xi_c + \rho\rightarrow\Xi_c+\pi+\pi{\Big ]}
= {3\over2} \times {\bf \Gamma\Big[}\Xi_c^+[1/2^-,2P] \rightarrow \Xi_c^0 + \pi^++\pi^0{\Big ]} \, ,
\\ &(c19)& {\bf \Gamma\Big[}\Xi_c[1/2^-,2P] \rightarrow \Sigma_c^* + \bar K^*\rightarrow\Sigma_c^{*}+\bar K+\pi{\Big ]}
= {9\over2} \times {\bf \Gamma\Big[}\Xi_c^+[1/2^-,2P] \rightarrow \Sigma_c^{*++} + K^-+ \pi^0{\Big ]} \, ,
\\ &(c20)& {\bf \Gamma\Big[}\Xi_c[1/2^-,2P] \rightarrow \Xi_c^{*}+ \rho\rightarrow\Xi_c^{*}+\pi+\pi {\Big ]}
={3\over2} \times {\bf \Gamma\Big[}\Xi_c^+[1/2^-,1P] \rightarrow \Xi_c^{*0} + \pi^0+\pi^+ {\Big ]}\, ,
\\ &(d1)& {\bf \Gamma\Big[}\Xi_c[3/2^-,1P] \rightarrow \Lambda_b + \bar K{\Big ]}
= {\bf \Gamma\Big[}\Xi_c^0[3/2^-,1P] \rightarrow\Lambda_c^+ +K^-{\Big ]} \, ,
\\ &(d2)&{\bf \Gamma\Big[}\Xi_c[3/2^-,1P] \rightarrow\Xi_c + \pi{\Big ]}
= {3\over2} \times {\bf \Gamma\Big[}\Xi_c^0[3/2^-,1P] \rightarrow \Xi_c^+ +\pi^-{\Big ]} \, ,
\\ &(d3)&{\bf \Gamma\Big[}\Xi_c[3/2^-,1P] \rightarrow \Sigma_c + \bar K{\Big ]}
= 3 \times {\bf \Gamma\Big[}\Xi_c^0[3/2^-,1P] \rightarrow \Sigma_c^+ +K^-{\Big ]} \, ,
\\ &(d4)& { \bf\Gamma\Big[}\Xi_c[3/2^-,1P] \rightarrow \Xi_c^{\prime}+\pi {\Big]}
= {3\over2}\times{\bf \Gamma\Big[}\Xi_c^0[3/2^-,1P]\rightarrow\Xi_c^{\prime+}+\pi^-{\Big]} \, ,
\\ &(d5)& {\bf \Gamma\Big[}\Xi_c[3/2^-,1P] \rightarrow \Sigma_c^* + \bar K{\Big ]}
= 3 \times {\bf \Gamma\Big[}\Xi_c^0[3/2^-,1P] \rightarrow \Sigma_c^{*+} + K^-{\Big ]} \, ,
\\ &(d6)& {\bf \Gamma\Big[}\Xi_c[3/2^-,1P] \rightarrow \Xi_c^* + \pi\rightarrow \Xi_c+\pi+\pi {\Big ]}
= {9 \over 2} \times {\bf \Gamma\Big[}\Xi_c^{*0}[3/2^-,1P] \rightarrow\Xi_c^{*+} + \pi^-\rightarrow\Xi_c^++\pi^0+\pi^-{\Big]} \, ,
\\ &(d7)& {\bf \Gamma\Big[}\Xi_c[3/2^-,1P] \rightarrow \Lambda_c + \bar{K}^*\rightarrow\Lambda_b+\bar K+\pi{\Big ]}
=3\times  {\bf \Gamma\Big[}\Xi_c^0[3/2^-,1P] \rightarrow \Lambda_c^+ +  K^- +\pi^0{\Big ]} \, ,
\\ &(d8)& {\bf \Gamma\Big[}\Xi_c[3/2^-,1P] \rightarrow \Xi_c + \rho\rightarrow\Xi_b+\pi+\pi{\Big ]}
= {3\over2} \times {\bf \Gamma\Big[}\Xi_c^0[3/2^-,1P] \rightarrow \Xi_c^+ + \pi^0+\pi^-{\Big ]} \, ,
\\ &(d9)& {\bf \Gamma\Big[}\Xi_c[3/2^-,1P] \rightarrow \Sigma_c^* + \bar K^*\rightarrow\Sigma_c^{*0}+\bar K+\pi{\Big ]}
= 9 \times {\bf \Gamma\Big[}\Xi_c^0[3/2^-,1P] \rightarrow \Sigma_c^{*+} + K^-+ \pi^0{\Big ]} \, ,
\\ &(d10)& {\bf \Gamma\Big[} \Xi_c[3/2^-,1P] \rightarrow \Xi_c^{*}+ \rho\rightarrow\Xi_c^{*}+\pi+\pi {\Big ]}
={3\over2} \times {\bf \Gamma\Big[}\Xi_c^0[3/2^-,1P] \rightarrow \Xi_c^{*+} + \pi^0+\pi^- {\Big ]}\, ,
\\ &(d11)& {\bf \Gamma\Big[}\Xi_c[3/2^-,2P] \rightarrow \Lambda_c + \bar K{\Big ]}
= {\bf \Gamma\Big[}\Xi_c^+[3/2^-,2P] \rightarrow\Lambda_c^+ +\bar{K}^0{\Big ]} \, ,
\\ &(d12)&{\bf \Gamma\Big[}\Xi_c[3/2^-,2P] \rightarrow\Xi_c + \pi{\Big ]}
= {3\over2} \times {\bf \Gamma\Big[}\Xi_c^+[3/2^-,2P] \rightarrow \Xi_c^0 +\pi^+{\Big ]} \, ,
\\ &(d13)&{\bf \Gamma\Big[}\Xi_c[3/2^-,2P] \rightarrow \Sigma_c + \bar K{\Big ]}
= {3\over2} \times {\bf \Gamma\Big[}\Xi_c^+[3/2^-,2P] \rightarrow \Sigma_c^{++} +K^-{\Big ]} \, ,
\\ &(d14)& { \bf\Gamma\Big[}\Xi_c[3/2^-,2P] \rightarrow \Xi_c^{\prime}+\pi {\Big]}
= {3\over2}\times{\bf \Gamma\Big[}\Xi_c^+[3/2^-,2P]\rightarrow\Xi_c^{\prime0}+\pi^+{\Big]} \, ,
\\ &(d15)& {\bf \Gamma\Big[}\Xi_c[3/2^-,2P] \rightarrow \Sigma_c^* + \bar K{\Big ]}
= {3\over2} \times {\bf \Gamma\Big[}\Xi_c^+[3/2^-,2P] \rightarrow \Sigma_c^{*++} + K^-{\Big ]} \, ,
\\ &(d16)& {\bf \Gamma\Big[}\Xi_b[3/2^-,2P] \rightarrow \Xi_c^* + \pi{\Big ]}
= {3 \over 2} \times {\bf \Gamma\Big[}\Xi_c^{*+}[3/2^-,2P] \rightarrow\Xi_c^{*0} + \pi^+{\Big]} \, ,
\\ &(d17)& {\bf \Gamma\Big[}\Xi_c[3/2^-,2P] \rightarrow \Lambda_c + \bar{K}^*\rightarrow\Lambda_b+\bar K+\pi{\Big ]}
=3\times  {\bf \Gamma\Big[}\Xi_c^+[3/2^-,2P] \rightarrow \Lambda_c^+ +  \bar{K}^0 +\pi^0{\Big ]} \, ,
\\ &(d18)& {\bf \Gamma\Big[}\Xi_c[3/2^-,2P] \rightarrow \Xi_c + \rho\rightarrow\Xi_b+\pi+\pi{\Big ]}
= {3\over2} \times {\bf \Gamma\Big[}\Xi_c^+[3/2^-,2P] \rightarrow \Xi_c^0 + \pi^0+\pi^+{\Big ]} \, ,
\\ &(d19)& {\bf \Gamma\Big[}\Xi_c[3/2^-,2P] \rightarrow \Sigma_c^* + \bar K^*\rightarrow\Sigma_c^{*}+\bar K+\pi{\Big ]}
= {9\over2} \times {\bf \Gamma\Big[}\Xi_c^+[3/2^-,1P] \rightarrow \Sigma_c^{*++} + K^-+ \pi^0{\Big ]} \, ,
\\ &(d20)& {\bf \Gamma\Big[} \Xi_c[3/2^-,2P] \rightarrow \Xi_c^{*}+ \rho\rightarrow\Xi_c^{*}+\pi+\pi {\Big ]}
={3\over2} \times {\bf \Gamma\Big[}\Xi_c^+[3/2^-,1P] \rightarrow \Xi_c^{*0} + \pi^0+\pi^+ {\Big ]}\, .
\label{eq:couple}
\end{eqnarray}
\end{widetext}
We shall calculate their partial decay widths through the following Lagrangians:
\begin{eqnarray}
&&\mathcal{L}^S_{X_c({1/2}^-) \rightarrow Y_c({1/2}^+) \mathcal P} = g {\bar X_c} Y_c \mathcal P \, ,
\\ &&\mathcal{L}^S_{X_c({3/2}^-) \rightarrow Y_c({3/2}^+) \mathcal P} = g {\bar X_{c\mu}}Y_c^{\mu} \mathcal P \, ,
\\ &&\mathcal{L}^S_{X_c({1/2}^-) \rightarrow Y_c({1/2}^+) V} = g {\bar X_c} \gamma_\mu \gamma_5 Y_c V^\mu \, ,
\\ &&\mathcal{L}^S_{X_c({1/2}^-) \rightarrow Y_c({3/2}^+) V} = g {\bar X_{c}} Y_{c}^{\mu} V_\mu \, ,
\\ &&\mathcal{L}^S_{X_c({3/2}^-) \rightarrow Y_c({1/2}^+) V} = g {\bar X_{c}^{\mu}} Y_{c} V_\mu \, ,
\\ &&\mathcal{L}^S_{X_c({3/2}^-) \rightarrow Y_c({3/2}^+) V} = g {\bar X_c}^{\nu} \gamma_\mu \gamma_5 Y_{c\nu} V^\mu \, .
\\ && \mathcal{L}^D_{X_c({1/2}^-) \rightarrow Y_c({3/2}^+) \mathcal P} = g {\bar X_c} \gamma_\mu \gamma_5 Y_{c\nu} \partial^{\mu} \partial^{\nu}\mathcal P \, ,
\\ && \mathcal{L}^D_{X_c({3/2}^-) \rightarrow Y_c({1/2}^+) \mathcal P} = g {\bar X_{c\mu}} \gamma_\nu \gamma_5 Y_{c} \partial^{\mu} \partial^{\nu}\mathcal P \, ,
\\ && \mathcal{L}^D_{X_c({3/2}^-) \rightarrow Y_c({3/2}^+) \mathcal P} = g {\bar X_{c\mu}} Y_{c\nu} \partial^{\mu} \partial^{\nu}\mathcal P \, .
\label{eq:lagrangians}
\end{eqnarray}
Here the superscripts $S$ and $D$ indicate the $S$- and $D$-wave decays, respectively; the fields $X_c^{(\mu)}$ and $Y_c^{(\mu)}$ represent the $P$-wave charmed baryons and ground-state charmed baryons, respectively; the fields $\mathcal{P}$ and $V^\mu$ denote the light pseudoscalar mesons and light vector mesons, respectively. To facilitate a comprehensive analysis of the aforementioned decay channels, we employ the pertinent parameters of ground-states, excited states, pseudoscalar, and vector mesons as follows:
\begin{itemize}
\item Relevant parameters of ground-state charmed baryons are
\begin{eqnarray}
\nonumber\Lambda_c(1/2^+)&:& M=2286.46~{\rm MeV}\, ,
\\ \nonumber \Xi_c(1/2^+)&:& M=2469.34~{\rm MeV}\, ,
\\ \nonumber \Sigma_c(1/2^+)&:& M=2453.54~{\rm MeV}\, ,
\\ \nonumber \Sigma^*_c(3/2^+)&:& M=2518.1~{\rm MeV}\, ,
\\ \nonumber \Xi^\prime_c(1/2^+)&:& M=2576.8~{\rm MeV}\, ,
\\ \nonumber \Xi^*_c(3/2^+)&:& M=2645.9~{\rm MeV}\, ,
\\ \nonumber \Omega_c(1/2^+)&:& M=2695.2~{\rm MeV}\, ,
\\ \nonumber \Omega^*_c(1/2^+)&:& M=2765.9~{\rm MeV}\, .
\end{eqnarray}

\item Relevant parameters of excited state charmed baryons are
\begin{eqnarray}
\nonumber \Lambda_c(1/2^-,1P) &:&m=2592.25~{\rm MeV}\, ,
\\ \nonumber \Lambda_c(3/2^-,1P)&:& m=2628.11~{\rm MeV}\, ,
\\ \nonumber \Xi_c(1/2^-,1P)&:& m= 2790.45~{\rm MeV}\, ,
\\ \nonumber \Xi_c(3/2^-,1P)&:& m=2818.1~{rm MeV}\, ,
\\ \nonumber \Lambda_c(1/2^-,2P)&:& m=2913.8~{\rm MeV}\, ,
\\ \nonumber \Lambda_c(3/2^-,2P)&:& m=2939.6~{\rm MeV}\, ,
\\ \nonumber \Xi_c(3/2^-,2P)&:& m=3122.9 ~{\rm MeV}\, ,
\\ \nonumber \Xi_c(1/2^-,2P)&:& m=m[\Xi_c(3/2^-,2P)]-\Delta m
\\ \nonumber &&~~~=(3122.9-18)~{\rm MeV}\, .
\end{eqnarray}

\item Relevant parameters of light pseudoscalar and vector mesons are
\begin{eqnarray}
\nonumber \pi(0^-)&:& m=138.04~{\rm MeV}\, ,
\\ \nonumber K(0^-) &:& m=495.65 ~{\rm MeV}\, ,
\\ \nonumber \rho(1^-)&:& m=775.21~{\rm MeV}\, ,
\\ \nonumber     &&\Gamma=148.2~{\rm MeV}\, ,
\\ \nonumber   &&g_{\rho\pi\pi}=5.94~{\rm GeV}^{-2}\, ,
\\ \nonumber K^{*}(1^-)&:& m=893.57 ~{\rm MeV}\, ,
\\ \nonumber &&\Gamma=49.1~{\rm MeV}
\\ \nonumber &&g_{K^* k\pi}=3.20~{\rm GeV}^{-2}\, .
\end{eqnarray}
\end{itemize}

We use the $\Lambda_c(1P,{1/2}^-)$ and $\Lambda_c(2P,{1/2}^-)$ from the $[\mathbf{\bar 3}_F, 1, 1, \rho]$ doublet as an example, and study their $S$-wave decays into the ground-state charmed baryon $\Sigma_c^{++}$ and the light pseudoscalar meson $\pi^-$. To do this we investigate the two-point correlation function:
\begin{eqnarray}
\nonumber\Pi(\omega_1, \omega_2,\omega^\prime)& =& \int d^4 x~e^{-i k \cdot x}~\langle 0 | J^\alpha_{1/2,-,\Lambda_c^-}(0) \bar J_{\Sigma_c^{++}}(x) | \pi^-(q) \rangle
\\  &=& {1+v\!\!\!\slash\over2} G^\alpha_{\Lambda_c^+({1\over2}^-) \rightarrow   \Sigma_c^{++}\pi^-} (\omega_1, \omega_2,\omega^\prime) \, .
\end{eqnarray}

At the hadron level we write
$G_{\Lambda_c^+({1\over2}^-) \rightarrow \Sigma_c^{++}\pi^-}$ as:
\begin{eqnarray}
&&G_{\Lambda_c^+({1\over2}^-) \rightarrow \Sigma_c^{++}\pi^-} (\omega_1, \omega_2,\omega^\prime_1)\\ \nonumber
&=& g_{\Lambda_c^+({1\over2}^-,1P) \rightarrow \Sigma_c^{++}\pi^-} { f_{\Lambda_c({1\over2}^-,1P)} f_{\Sigma_c^{++}} \over (\bar \Lambda_{\Lambda_c({1\over2}^-,1P)} - \omega_1) (\bar \Lambda_{\Sigma_c^{++}} - \omega^\prime)}\\
\nonumber  &+& g_{\Lambda_c^+({1\over2}^-,1P) \rightarrow \Sigma_c^{++}\pi^-} {  f_{\Lambda_c({1\over2}^-,2P)} f_{\Sigma_c^{++}} \over (\bar \Lambda_{\Lambda_c({1\over2}^-,2P)} - \omega_2) (\bar \Lambda_{\Sigma_c^{++}} - \omega^\prime)}\cdots \, ,
\label{G0C}
\end{eqnarray}
where $\cdots$ contains other possible amplitudes.

At the quark-gluon level we calculate $G^\alpha_{\Lambda_c^+[{1\over2}^-] \rightarrow \Sigma_c^{++}\pi^-}$ using the method of operator product expansion (OPE). Then we perform the Borel transformation to both  hadron level and quark-gluon level:
\begin{widetext}
\begin{eqnarray}
\label{eq:g}
\nonumber && g^S_{\Lambda_c^+({1\over2}^-,1P)\rightarrow \Sigma_c^{++}\pi^-} f_{\Lambda_c({1\over2}^-,1P)} f_{\Sigma_c^{++}} e^{- {\bar \Lambda_{\Lambda_c({1\over2}^-,1P)} \over T_1}} e^{ - {\bar \Lambda_{\Sigma_c^{++}} \over T^\prime}}+g^S_{\Lambda_c^-({1\over2}^-,2P) \rightarrow \Sigma_c^{++}\pi^-} f_{\Lambda_c({1\over2}^-,2P)} f_{\Sigma_c^{++}} e^{- {\bar \Lambda_{\Lambda_c({1\over2}^-,2P)} \over T_2}} e^{ - {\bar \Lambda_{\Sigma_c^{++}} \over T^\prime}}
\\ \nonumber &=&4\sqrt{2} \times \Big (-\int_0^{1\over2}du_1\frac{3f_\pi}{2\pi^2}T^4f_3({\omega_c^\prime\over T})\psi_{4;\pi}(u_1)-\frac{3f_\pi}{2\pi^2}T^6f_5({\omega_c^\prime\over T})\frac{\partial}{\partial u_0}\phi_{2;\pi}(u_0)|_{u_0={1\over2}}+\frac{f_\pi}{2\pi^2}T^6f_5({\omega_c^\prime\over T})\frac{\partial^2}{\partial u_0^2}u_0\phi_{2;\pi}(u_0)|_{u_0={1\over2}}
\\ \nonumber &&
~~~~~+\frac{3f_\pi}{32\pi^2}T^4f_3({\omega_c^\prime\over T})\frac{\partial}{\partial u_0}\phi_{4;\pi}(u_0)|_{u_0={1\over2}}
+\frac{f_\pi m_\pi^2 m_s}{8(m_u+m_d)\pi^2}T^4f_3({\omega_c^\prime\over T})\frac{\partial}{\partial u_0}\phi^\sigma_{3;\pi}(u_0)|_{u_0={1\over2}}
\\ \nonumber &&~~~~~+\frac{f_\pi}{32\pi^2}T^4f_3({\omega_c^\prime\over T})\frac{\partial^2}{\partial u_0^2}u_0\phi_{4;\pi}(u_0)|_{u_0={1\over2}}
-\frac{f_\pi m_\pi^2 m_s}{24(m_u+m_d)\pi^2}T^4f_3({\omega_c^\prime\over T})\frac{\partial^2}{\partial u_0^2}u_0\phi^\sigma_{3;\pi}(u_0)|_{u_0={1\over2}}
\\ \nonumber &&
~~~~~-\frac{ f_\pi m_s}{16}\langle \bar q q\rangle  T^2 f_1({\omega_c^\prime\over T})\frac{\partial}{\partial u_0} \phi_{2;\pi}(u_0)|_{u_0={1\over2}}
-\frac{f_\pi m_\pi^2}{24(m_u+m_d)}\langle \bar q q\rangle T^2 f_1({\omega_c^\prime\over T})\frac{\partial}{\partial u_0}\phi^\sigma_{3;\pi}(u_0)|_{u_0={1\over2}}
\\ \nonumber &&
~~~~~-\int_0^{1\over2} du_1\frac{f_\pi m_s}{16}\langle\bar q q\rangle\psi_{4;\pi}(u_1)
+\frac{f_\pi m_\pi^2}{72(m_u+m_d)}\langle\bar q q\rangle T^2f_1({\omega_c^\prime\over T})\frac{\partial^2}{\partial u_0^2}u_0\phi^\sigma_{3;\pi}(u_0)|_{u_0={1\over2}}
\\ \nonumber &&
~~~~~-\frac{f_\pi m_s}{48}\langle\bar q q\rangle T^2f_1({\omega_c^\prime\over T})\frac{\partial^2}{\partial u_0^2} u_0\phi_{2;\pi}(u_0)|_{u_0={1\over2}}+\frac{f_\pi m_\pi^2}{384(m_u+m_d)}\langle g_s \bar q\sigma G q\rangle\frac{\partial}{\partial u_0}\phi^\sigma_{3;\pi}(u_0)|_{u_0={1\over2}}
\\ \nonumber &&
~~~~~+\frac{f_\pi m_s}{256}\langle\bar q q\rangle \frac{\partial}{\partial u_0}\phi_{4;\pi}(u_0)|_{u_0={1\over2}}
-\frac{f_\pi m_\pi^2}{1152(m_u+m_d)}\langle g_s\bar q \sigma G q\rangle\frac{\partial^2}{\partial u_0^2}u_0\phi^\sigma_{3;\pi}(u_0)|_{u_0={1\over2}}
\\ \nonumber &&
~~~~~+\frac{f_\pi m_s}{768}\langle \bar q q\rangle\frac{\partial^2}{\partial u_0^2}u_0\phi_{4;\pi}(u_0)|_{u_0={1\over2}}\Big )
\\ \nonumber &&
-{\sqrt{2}\over2}\times \Big(-\frac{f_\pi}{4\pi^2 u_0^2}T^4f_3({\omega_c^\prime\over T})\int_0^{1\over 2}d\alpha_2\int_{{1\over2}-\alpha_2}^{1-\alpha_2}(-\frac{u_0^2}{\alpha_3}\frac{\partial^2}{\partial \alpha_3^2}\alpha_3\Phi_{4;\pi}(\underline{\alpha})-\frac{u_0\alpha_2}{\alpha_3}\frac{\partial^2}{\partial \alpha_3^3}\Phi_{4;\pi}(\underline{\alpha})-\frac{u_0}{2\alpha_3}\frac{\partial^2}{\partial\alpha_3^2}\alpha_3\Phi_{4;\pi}(\underline{\alpha})
\\ \nonumber &&
~~~~~-\frac{u_0}{2\alpha_3}\frac{\partial^2}{\partial\alpha_3^2}\alpha_3\widetilde\Phi_{4;\pi}(\underline{\alpha})+\frac{u_0}{\alpha_3}\frac{\partial^2}{\partial\alpha_3^2}\Phi_{4;\pi}(\underline{\alpha})-\frac{\alpha_2}{2\alpha_3}\frac{\partial^2}{\partial\alpha_3^2}\Phi_{4;\pi}(\underline{\alpha})-\frac{\alpha_2}{2\alpha_3}\frac{\partial^2}{\partial\alpha_3^2}\widetilde\Phi_{4;\pi}(\underline{\alpha})+\frac{1}{2\alpha_3}\frac{\partial^2}{\partial\alpha_3^2}\Phi_{4;\pi}(\underline{\alpha})\\ 
\nonumber &&~~~~~+\frac{1}{2\alpha_3}\frac{\partial^2}{\partial\alpha_3^2}\widetilde\Phi_{4;\pi}(\underline{\alpha}))
-\frac{ f_\pi}{2\pi^2 u_0}T^4f_3({\omega_c^\prime\over T}) \int_0^{1 \over 2} d\alpha_2 \int_{{1 \over 2}-\alpha_2}^{1-\alpha_2} d\alpha_3(\frac{ u_0}{2\alpha_3}\frac{\partial}{\partial\alpha_3}\Phi_{4;\pi}(\underline{\alpha})+\frac{u_0}{\alpha_3} \frac{\partial}{\partial\alpha_3}\Psi_{4;\pi}(\underline{\alpha})\\
\nonumber &&~~~~~+\frac{3 }{4\alpha_3}\frac{\partial}{\partial\alpha_3}\Phi_{4;\pi}(\underline{\alpha})
+\frac{1}{4\alpha_3}\frac{\partial}{\partial\alpha_3}\widetilde\Phi_{4;\pi}(\underline{\alpha})-\frac{1}{2\alpha_3}\frac{\partial}{\partial\alpha_3}\Psi_{4;\pi}(\underline{\alpha}) +\frac{1}{2\alpha_3}\frac{\partial}{\partial\alpha_3i}\widetilde\Psi_{4;\pi}(\underline{\alpha}))\Big ) \, .
\end{eqnarray}
\end{widetext}
In the above expressions, $f_n(x) \equiv 1 - e^{-x} \sum_{k=0}^n {x^k \over k!}$; the parameters $\omega_1$, $\omega_2$, and $\omega^\prime$ are transformed to be $T_1$, $T_2$, and $T^\prime$ respectively; we choose $T_1 = T_2 = T^\prime =2T$ so that $u_0 = {T_1 \over T_1 + T^\prime} = {1\over2}$; we choose $\omega_c = 1.20$~GeV to be the average threshold value of the $\Lambda_c^+({1/2}^-,1P)$ and $\Sigma_c^{++}(1/2^+)$ mass sum rules; we choose $\omega_c = 1.47$~GeV to be the average threshold value of the $\Lambda_c^+({1/2}^-,2P)$ and $\Sigma_c^{++}(1/2^+)$ mass sum rules; we choose $0.260~\rm{GeV}<T<0.266~\rm{GeV}$ to be the Borel window of the $\Lambda_c^+({1/2}^-,1P/2P)$ mass sum rule. The light-cone distribution amplitudes contained in the above sum rule expressions can be found in Refs.~\cite{Ball:1998je,Ball:2006wn,Ball:2004rg,Ball:1998kk,Ball:1998sk,Ball:1998ff,Ball:2007rt,Ball:2007zt}.

\begin{figure*}[hbt]
\begin{center}
\subfigure[]{
\scalebox{0.5}{\includegraphics{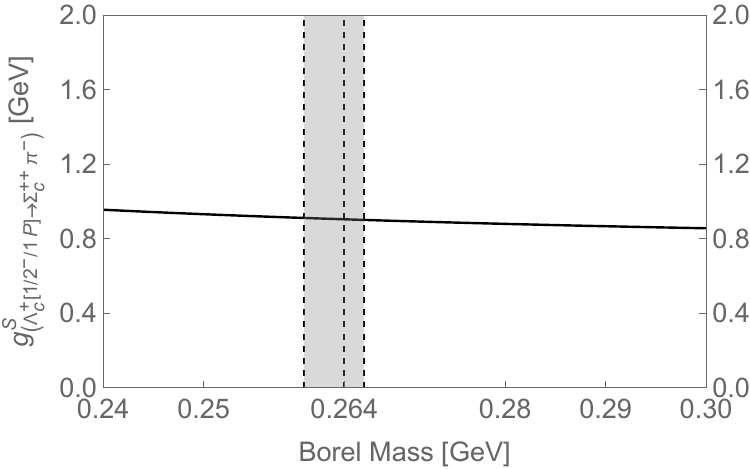}}}~~~~
\subfigure[]{
\scalebox{0.5}{\includegraphics{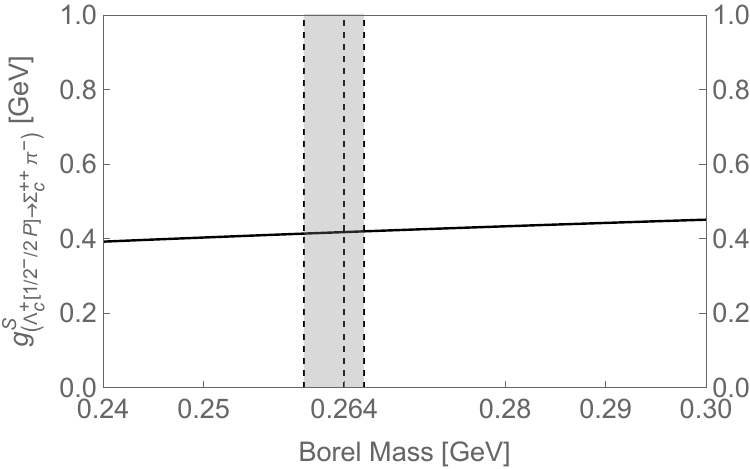}}}~~~~
\caption{Coupling constants as functions of the Borel mass $T$: (a) $g^S_{\Lambda_c[{1\over2}^-/1P]\to\Sigma_c^{++}\pi^-}$ and (b) $g^S_{\Lambda_c[{1\over2}^-/2P]\to\Sigma_c^{++}\pi^-}$.}
\label{fig:g}
\end{center}
\end{figure*}

We extract the coupling constant from Eq.~(\ref{eq:g}) to be:
\begin{eqnarray}
g^S_{\Lambda_c^+[{1\over2}^-,1P] \rightarrow \Sigma_c^{++}\pi^-} &=& 0.90^{+2.39}_{-0.90} \, ,
\\ \nonumber g^S_{\Lambda_c^+[{1\over2}^-,2P] \rightarrow \Sigma_c^{++}\pi^-} &=& 0.42^{+0.62}_{-0.42}\, ,
\end{eqnarray}
where the uncertainties are due to $T$, $\omega_c$, $\omega_c^\prime$, parameters of $\Lambda_c(1/2^+,1P)$, parameters of $\Sigma_c^{++}({1/2}^+)$, and various QCD parameters given in Eq.~(\ref{eq:condensates}). We depict $g^S_{\Lambda_c^-[{1\over2}^-,1P/2P] \rightarrow \Sigma_c^{++}\pi^0}$ in Fig.~\ref{fig:g} as a function of the Borel mass $T$. Its dependence on $T$ is weak inside the Borel window $0.260$~GeV$<T<0.266$~GeV.

The $S$-wave decay widths of $P$-wave charmed baryons can be calculated through the Lagrangians expressed in Eq.~(\ref{eq:lagrangians}):
\begin{itemize}
\item Since the $\Lambda_c(1/2^-,1P)$ is below the $\Sigma_c \pi$ threshold but above the $\Lambda_c \pi \pi$ threshold, we use the following formula for the three-body $S$-wave decay process $\Lambda_c(1/2^-,1P)\to \Sigma_c+\pi \to \Lambda_c + \pi + \pi$:
\begin{eqnarray}
&&\Gamma\left(\Lambda_c(1/2^-,1P)\to\Sigma_c+\pi\right)
\\ \nonumber &=& \Gamma\left(0\to 2+1\right)+2\times\Gamma\left(0\to 6+3\to 5+4+3\right)
\\ \nonumber &\equiv&\Gamma \left(\Lambda_c^+(1/2^-,1P)\to\Sigma_c^++\pi^0\right)\\
\nonumber &+&2 \times\Gamma \left(\Lambda_c^+(1/2^-,1P)\to\Sigma_c^{++}+\pi^-\to \Lambda_c^++\pi^++\pi^-\right)
\\ \nonumber &=& {|\vec{p_1}|\over 8\pi m_0^2}\times g_{0\to 2+1}^2\times {1\over2} Tr[(p_0\!\!\!\slash+m_0)(p_2\!\!\!\slash+m_2)]\\
\nonumber &+& 2 \times {1\over (2\pi)^3}\times {1\over 32 m_0^3}\times g_{0\to 6+3}^2\times g_{6\to 5+4}^2\times \int dm_{34} dm_{45}
\\ \nonumber &\times& {1\over2}Tr[(p_5\!\!\!\slash+m_5)\gamma_\nu^\prime \gamma_5(p_6\!\!\!\slash+m_6)(p_0\!\!\!\slash +m_0)(p_6\!\!\!\slash +m_6)\gamma_\mu\gamma_5]
\\ \nonumber &\times& {1\over |p_6^2-m_6^2+i m_6\Gamma_6|^2}\times p_{4,\mu}p_{4,\mu^\prime}\, .
\end{eqnarray}

\item Since the $\Lambda_c(1/2^-,2P)$ is above the $\Sigma_c \pi$ threshold, we use the following formula for the two-body $S$-wave decay process $\Lambda_c(1/2^-,2P)\to \Sigma_c+\pi$:
\begin{eqnarray}
&&\Gamma\left(\Lambda_c(1/2^-,2P)\to\Sigma_c+\pi\right)\\
\nonumber &=& 3\times\Gamma\left(0\to 6+3\right)\\
\nonumber &\equiv&3 \times\Gamma \left(\Lambda_c^+(1/2^-,2P)\to\Sigma_c^{++}+\pi^-\right)\\
\nonumber &=& {|\vec{p_3}|\over 8\pi m_0^2}\times g_{0\to 6+3}^2 \times {1\over2}Tr[(p_0\!\!\!\slash+m_0)(p_6\!\!\!\slash+m_6)] \, ,
\end{eqnarray}
\end{itemize}
where $0$ denotes the initial $\Lambda_c(1/2^-,1P/2P)$; $1$ and $2$ denote the finial states $\pi^0$ and $\Sigma_c^+(1/2^+)$, respectively; $3$, $4$, and $5$ denote the finial states $\pi^-$, $\pi^+$, and $\Lambda_c(1/2^+)$, respectively; $6$ denotes the middle (finial) state $\Sigma_c^{++}(1/2^+)$.

We calculate their partial decay widths to be
\begin{eqnarray}
\Gamma^S_{\Lambda_c^+[{1\over2}^-,1P] \rightarrow \Sigma_c\pi }&=& 4.9^{+60.0}_{-~4.9}{\rm~MeV} \, ,
\\ \nonumber \Gamma^S_{\Lambda_c^+[{1\over2}^-,2P] \rightarrow \Sigma_c\pi}&=& 29^{+148}_{-~29}{\rm~MeV} \, .
\end{eqnarray}
Similarly, we study the other decay channels for the charmed baryons belonging to the $[\mathbf{\bar 3}_F, 1, 1, \rho]$ and $[\mathbf{\bar 3}_F, 1, 0, \lambda]$ doublets.
Their partial decay widths are evaluated and summarized in Table~\ref{tab:decayc311rho} and Table~\ref{tab:decayc310lambda}.

\section{Discussions and summary}\label{sec:summary}
%

In this paper we investigate the $\Lambda_c(2910)^+$, $\Lambda_c(2940)^+$, and $\Xi_c(3123)^+$ as potential $2P$-wave charmed baryons with $J^P=1/2^-$ and $3/2^-$, all of which belong to the $[\mathbf{\bar 3}_F,1,1,\rho]$ doublet. This doublet contains four $P$-wave charmed baryons, namely $\Lambda_c(1/2^-)$, $\Lambda_c(3/2^-)$, $\Xi_c(1/2^-)$, and $\Xi_c(3/2^-)$. We make the assumption that the current $J_{\mathbf{\bar 3}_F,1,1,\rho}$ can simultaneously couple to $1P$- and $2P$-wave charmed baryons, consequently leading to the possible production of $1P$ and $2P$-wave excitations such as $\Lambda_c(2595)^+$, $\Lambda_c(2625)^+$, $\Xi_c(2790)^+$, $\Xi_c(2815)^+$, $\Lambda_c(2910)^+$, $\Lambda_c(2940)^+$, and $\Xi_c(3123)^+$. The masses of these baryons are computed utilizing the QCD sum rule method within the framework of heavy quark effective theory.  Additionally, we analyze their decays into ground-state charmed baryons as well as the light pseudoscalar mesons $\pi/K$ and vector mesons $\rho/K^*$ using the light-cone sum rule method for $S$- and $D$-wave decays. Based on these results, we derive the masses and total widths of the $P$-wave charmed baryons, facilitating a comprehensive understanding of their properties as a whole:
\begin{eqnarray}
\nonumber M_{\Lambda_c({1\over2}^-,1P)}&=&2.61\pm 0.07~\mbox{GeV}\, ,
\\ \nonumber \Gamma_{\Lambda_c({1\over2}^-,1P)}&=& 5^{+60}_{-~5} ~\mbox{MeV}\, ,
\\ \nonumber M_{\Lambda_c({3\over2}^-,1P)}&=& 2.63\pm 0.07~\mbox{GeV}\, ,
\\ \nonumber \Gamma_{\Lambda_c({3\over2}^-,1P)}&=& 0.2^{+2.4}_{-0.2}~\mbox{MeV}\, ,
\\ M_{\Lambda_c({3\over2}^-,1P)}-M_{\Lambda_c({1\over2}^-,1P)}&=&24\pm 9~\mbox{MeV}\, ,
\\ \nonumber M_{\Lambda_c({1\over2}^-,2P)}&=&3.05\pm 0.12~\mbox{GeV}\, ,
\\ \nonumber \Gamma_{\Lambda_c({1\over2}^-,2P)}&=& 35^{+148}_{-~29}~\mbox{MeV} \, ,
\\ \nonumber M_{\Lambda_c({3\over2}^-,2P)}&=& 3.07\pm 0.11~\mbox{GeV}\, ,
\\ \nonumber \Gamma_{\Lambda_c({3\over2}^-,2P)}&=& 29^{+140}_{-~26}~\mbox{MeV}\, ,
\\ M_{\Lambda_c({3\over2}^-,2P)}-M_{\Lambda_c({1\over2}^-,2P)}&=&18\pm 7~\mbox{MeV}\, ,
\\ \nonumber M_{\Xi_c({1\over2}^-,1P)}&=& 2.78\pm0.14~\mbox{GeV}\, ,
\\ \nonumber \Gamma_{\Xi_c({1\over2}^-,1P)}&=& 3^{+37}_{-~3}~\mbox{MeV}\, ,
\\ \nonumber M_{\Xi_c({3\over2}^-,1P)}&=& 2.81\pm0.14~\mbox{GeV}\, ,
\\ \nonumber \Gamma_{\Xi_c({3\over2}^-,1P)}&=& 0.6^{+18.9}_{-~0.6}~\mbox{MeV}\, ,
\\ \nonumber  M_{\Xi_c({3\over2}^-,1P)}-M_{\Xi_c(1P/{1\over2}^-)}&=& 29\pm 10~\mbox{MeV}\, ,
\\ \nonumber M_{\Xi_c({1\over2}^-,2P)}&=& 3.16\pm0.12~\mbox{GeV}\, ,
\\ \nonumber \Gamma_{\Xi_c({1\over2}^-,2P)}&=& 31^{+170}_{-~27}~\mbox{MeV}\, ,
\\ \nonumber M_{\Xi_c({3\over}^-,2P)}&=&3.18\pm0.12~\mbox{GeV}\, ,
\\ \nonumber \Gamma_{\Xi_c({3\over2}^-,2P)}&=&24^{+131}_{-~20}~\mbox{MeV}\, ,
\\ \nonumber  M_{\Xi_c({3\over2}^-,2P)}-M_{\Xi_c({1\over2}^-,2P)}&=& 18\pm 7~\mbox{MeV}\, .
\end{eqnarray}

Our results suggest that the $\Lambda_c(2910)^+$, $\Lambda_c(2940)^+$, and $\Xi_c(3123)^+$ can be suitably explained as the $2P$-wave charmed baryon of $J^P=1/2^-$ and $3/2^-$, belonging to the $[\mathbf{\bar 3}_F,1,1,\rho]$ doublet. However, the expected masses of $\Lambda_c(2910)$ and $\Lambda_c(2940)$ surpass experimental measurements by approximately $130$ MeV, although consistent with the results obtained in the quark models, which suggests that their masses are expected to exceed $3$ GeV~\cite{Ebert:2011kk,Chen:2014nyo,Chen:2016iyi}. Additionally, our analysis of their decay properties is detailed in Table~\ref{tab:decayc311rho}. For the $\Lambda_c(2910)$ and $\Lambda_c(2940)$, in fact, they can not only decay into charmed baryons and light mesons, but also decay into charmed mesons and light baryons. In this work, we only investigate the former so the total width of $\Lambda_c(2P,{1\over2}^-)$ is about $20$ MeV smaller than the measurement of $\Lambda_c(2910)^+$. Accounting for its $S$-wave decay into the $D^0p$ final state would result in a greater width. Similarly, the $\Lambda_c({3\over2}^-,2P)$ can also decay into $D^0p$ through a $D$-wave decay process, resulting in a slight increase in its total width, which is consistent with the experimental result of $\Lambda_c(2940)$ within the margin of error. The $\Xi_c(3123)^+$ could be interpreted as the $\Lambda_c(2940)^+$'s charmed-strange partner state with $J^P=3/2^-$. Furthermore, a $2P$-wave charmed-strange baryon, the $\Xi_b({1\over2}^-,2P)$, with a mass  $\Delta M=18\pm7$~MeV smaller than the $\Xi_c(3123)^+$, remains undiscovered. We propose to explore its presence in the $\Xi_c({1\over2}^-,2P)\to \Sigma_c K$ decay channel.

Apart from the $[\mathbf{\bar 3}_F, 1, 1, \rho]$ doublet, we also investigate $1P$ and $2P$-wave charmed baryons belonging to the $[\mathbf{\bar 3}_F, 1, 0, \lambda]$ doublet. The results, summarized in Appendix~\ref{sec:charmedsumrule}, indicate that the masses of $\Lambda_c(2910)^+$, $\Lambda_c(2940)^+$, $\Xi_c(2790)$, $\Xi_c(2815)$, and $\Xi_c(3123)^+$ alone cannot distinguish whether they belong to the $[\mathbf{\bar 3}_F, 1, 1, \rho]$ doublet or the $[\mathbf{\bar 3}_F,1,0,\lambda]$ doublet. However, considering their decay properties, the $1P$ and $2P$-wave charmed baryons belonging to the $[\mathbf{\bar 3}_F, 1, 0, \lambda]$ doublet exhibit wide widths, rendering them unobservable experimentally due to their extremely large coupling constants at the vertexes of their $S$-wave decay processes. Furthermore, our QCD sum rule results for the $\Lambda_c(2595)$, $\Lambda_c(2625)$, $\Lambda_c(2910)$, $\Lambda_c(2940)$, $\Xi_c(2790)$, $\Xi_c(2815)$, and $\Xi_c(3123)$ appear inconsistent with the quark model's expectations, where the $\rho$-mode multiplet $[\mathbf{\bar 3}_F, 1, 1, \rho]$ is anticipated to be higher in mass than the $\lambda$-mode multiplet $[\mathbf{\bar 3}_F, 1, 0, \lambda]$. This discrepancy could be attributed to considerable uncertainties in the mass differences between different multiplets within our framework. Furthermore, our previous study~\cite{Yang:2022oog} suggest that the $\Lambda_c(2595)$, $\Lambda_c(2635)$, $\Xi_c(2790)$, and $\Xi_c(2815)$ could potentially be explained as singly heavy baryons belonging to $[\mathbf{\bar 3}_F, 1, 1, \rho]$ doublet. Thus, it is crucial to verify the existence of $\rho$-mode heavy baryons, and we recommend to conduct an investigation into the existence of the $\rho$-mode. For instance, these exist four excited $\Omega_c$ baryons that are more likely to belong to the $\lambda$-mode multiplets in five $\Omega_c$ baryons observed by LHCb~\cite{LHCb:2017uwr}, while the possible assignment for the rest particle is either the radial $2S$-wave excitation or the orbital $1P$-wave excitation of the $\rho$-mode~\cite{Yoshida:2015tia,Nagahiro:2016nsx}.

Summarizing the above results, we study the $1P$ and $2P$-wave charmed baryons using the methods of QCD sum rules and light-cone sum rules within the framework of heavy quark effective theory. Our results suggest that the $\Lambda_c(2910)^+$, $\Lambda_c(2940)^+$, and $\Xi_c(3123)^+$ can be well interpreted as the $2P$-wave charmed baryon of $J^P=1/2^-$ and $3/2^-$, belonging to the $SU(3)$ flavor $\mathbf{\bar 3}_F$ representation. The $\Xi_c(3123)^+$ has a partner state of $J^P=1/2^-$, labeled as $\Xi_c(1/2^-,2P)$, whose mass and width are calculated to be $m_{\Xi_c(1/2^-,2P)} - m_{\Xi_c(3123)^+} = -18\pm{7}$~MeV and $\Gamma_{\Xi_c(2P,1/2^-)}=31^{+170}_{-~27}$~MeV, with $m_{\Xi_c(3123)^+} = 3122.9\pm{1.3}$~MeV. We suggest to search for this state in the $\Xi_c(2P,1/2^-)\to \Sigma_c K$ decay channel.

\begin{table*}[hbt]
\begin{center}
\renewcommand{\arraystretch}{1.5}
\caption{Decay properties of the $1P$ and $2P$-wave charmed baryons belonging to the $[\mathbf{\bar 3}_F, 1, 1, \rho]$ doublets, calculated using the method of light-cone sum rules within the framework of heavy quark effective theory. Experimental candidates are given in the last column.}
\setlength{\tabcolsep}{0.1mm}{
\begin{tabular}{ c | c | c | c | c | c | c | c}
\hline\hline
Baryon & ~~~~~~Mass~~~~~~ & Difference & \multirow{2}{*}{~~~~~~~~~Decay channels~~~~~~~~~}  & ~$S$-wave width~  & ~$D$-wave width~ & ~Total width~ & \multirow{2}{*}{Candidate}
\\  ($j^P$) & ({GeV})& ({MeV}) & & ({MeV}) & ({MeV}) & ({MeV})
\\ \hline\hline
\multirow{3}{*}{$\Lambda_c(1P,{1\over2}^-)$}& \multirow{3}{*}{$2.61\pm 0.07$} &\multirow{6}{*}{$24\pm 9$}&$\Lambda_c({1\over2}^-)\to \Sigma_c\pi\to\Lambda_c\pi\pi$&$4.9^{+60.0}_{-~4.9}$&--&\multirow{3}{*}{$5^{+60}_{-~5}$}&\multirow{3}{*}{$\Lambda_c(2595)$}
\\ \cline{4-6}
&&&$\Lambda_c({1\over2}^-)\to\Sigma_c^*\pi\to\Lambda_c\pi\pi$&--&$4\times10^{-8}$&&
\\ \cline{4-6}
&&&$\Lambda_c({1\over2}^-)\to \Lambda_c\rho\to \Lambda_c\pi\pi$& \multicolumn{2}{c|}{$1\times10^{-3}$ }&&
\\ \cline{1-2} \cline{4-8}
\multirow{3}{*}{$\Lambda_c(1P,{3\over2}^-)$}&\multirow{3}{*}{$2.64\pm 0.07$}&&$\Lambda_c({3\over2}^-)\to \Sigma_c \pi$ & -- & $4\times10^{-3}$&\multirow{3}{*}{$0.2^{+2.4}_{-0.2}$}&\multirow{3}{*}{$\Lambda_c(2625)$}
\\ \cline{4-6}
&&&$\Lambda_c({3\over2}^-)\to\Sigma_c^*\pi\to\Lambda_c\pi\pi$&$0.20^{+2.44}_{-0.20}$&$6\times10^{-7}$&&
\\ \cline{4-6}
&&&$\Lambda_c({3\over2}^-)\to \Lambda_c\rho\to \Lambda_c\pi\pi$&\multicolumn{2}{c|}{$0.02$}&&
\\ \cline{1-8}
\multirow{5}{*}{$\Lambda_c(2P,{1\over2}^-)$}& \multirow{5}{*}{$3.05\pm 0.12$} &\multirow{10}{*}{$18\pm 7$}&$\Lambda_c({1\over2}^-)\to \Sigma_c\pi$&$29^{+148}_{-~29}$&--&\multirow{5}{*}{$35^{+148}_{-~29}$}&\multirow{5}{*}{$\Lambda_c(2910)$}
\\ \cline{4-6}
&&&$\Lambda_c({1\over2}^-)\to\Sigma_c^*\pi$&--&$0.11^{+0.26}_{-0.10}$&&
\\ \cline{4-6}
&&&$\Lambda_c({1\over2}^-)\to \Lambda_c\rho\to \Lambda_c\pi\pi$& \multicolumn{2}{c|}{$0.47^{+1.51}_{-0.47}$}&&
\\ \cline{4-6}
&&&$\Lambda_c({1\over2}^-)\to \Sigma_c\rho\to \Sigma_c\pi\pi$& \multicolumn{2}{c|}{$5.0^{+9.6}_{-4.1}$ }&
\\ \cline{4-6}
&&&$\Lambda_c({1\over2}^-)\to \Sigma^*_c\rho\to \Sigma^*_c\pi\pi$& \multicolumn{2}{c|}{$0.26^{+0.50}_{-0.22}$ }&
\\ \cline{1-2}\cline{4-8}
\multirow{5}{*}{$\Lambda_c(2P,{3\over2}^-)$}&\multirow{5}{*}{$3.07\pm 0.11$}&&$\Lambda_c({3\over2}^-)\to \Sigma_c \pi$ & -- & $0.19^{+0.45}_{-0.17}$&\multirow{5}{*}{$29^{+140}_{-~26}$}&\multirow{5}{*}{$\Lambda_c(2940)$}
\\ \cline{4-6}
&&&$\Lambda_c({3\over2}^-)\to\Sigma_c^*\pi$&$26^{+140}_{-~26}$&$0.02$&&
\\ \cline{4-6}
&&&$\Lambda_c({3\over2}^-)\to \Lambda_c\rho\to \Lambda_c\pi\pi$&\multicolumn{2}{c|}{$0.53^{+1.59}_{-0.53}$}&&
\\ \cline{4-6}
&&&$\Lambda_c({3\over2}^-)\to \Sigma_c\rho\to \Sigma_c\pi\pi$&\multicolumn{2}{c|}{$1.4^{+2.6}_{-1.1}$}&
\\ \cline{4-6}
&&&$\Lambda_c({3\over2}^-)\to \Sigma_c^*\rho\to \Sigma_c^*\pi\pi$&\multicolumn{2}{c|}{$0.76^{+~1.47}_{-0.63}$}&
\\ \hline\hline
\multirow{3}{*}{$\Xi_c(1P,{1\over2}^-)$}&\multirow{3}{*}{$2.78\pm 0.14$}&\multirow{6}{*}{$29\pm 10$}&$\Xi_c({1\over2}^-)\to \Xi_c^{\prime}\pi$&$3.1^{+36.5}_{-~3.1}$&--&\multirow{3}{*}{$3^{+37}_{-~3}$}&\multirow{3}{*}{$\Xi_c(2790)$}
\\ \cline{4-6}
&&&$\Xi_c({1\over2}^-)\to\Xi_c^*\pi\to\Xi_c\pi\pi$&--&$2\times10^{-5}$&&
\\ \cline{4-6}
&&&$\Xi_c({1\over2}^-)\to\Xi_c\rho\to\Xi_c\pi\pi$&\multicolumn{2}{c|}{$3\times10^{-3}$}&&
\\ \cline{1-2} \cline{4-8}
\multirow{3}{*}{$\Xi_c(1P,{3\over2}^-)$}&\multirow{3}{*}{$2.81\pm 0.14$}&&$\Xi_c({3\over2}^-)\to\Xi_ c^{*}\pi\to\Xi_c\pi\pi$&$0.54^{+18.85}_{-~0.54}$&$1\times10^{-4}$&\multirow{3}{*}{$0.6^{+18.9}_{-~0.6}$}&\multirow{3}{*}{$\Xi_c(2815)$}
\\ \cline{4-6}
&&&$\Xi_c({3\over2}^-)\to\Xi_c^{\prime}\pi$&--&$0.05$&&
\\ \cline{4-6}
&&&$\Xi_c({3\over2}^-)\to \Xi_c\rho\to\Xi_c\pi\pi$&\multicolumn{2}{c|}{$0.02$}&&
\\ \hline
\multirow{9}{*}{$\Xi_c(2P,{1\over2}^-)$}&\multirow{9}{*}{$3.16\pm 0.12$}&\multirow{18}{*}{$18\pm 7$}&$\Xi_c({1\over2}^-)\to \Xi_c^{\prime}\pi$&$2.7^{+22.5}_{-~2.7}$&--&\multirow{9}{*}{$31^{+170}_{-~27}$}&\multirow{9}{*}{--}
\\ \cline{4-6}
&&&$\Xi_c({1\over2}^-)\to\Sigma_c K$&$27^{+168}_{-~27}$&--&&
\\ \cline{4-6}
&&&$\Xi_c({1\over2}^-)\to\Xi_c^*\pi$&--&$0.02$&&
\\ \cline{4-6}
&&&$\Xi_c({1\over2}^-)\to\Sigma_c^* K$&--&$3\times 10^{-3}$&&
\\ \cline{4-6}
&&&$\Xi_c({1\over2}^-)\to\Xi_c\rho\to\Xi_c\pi\pi$&\multicolumn{2}{c|}{$0.11^{+0.71}_{-0.11}$}&&
\\ \cline{4-6}
&&&$\Xi_c({1\over2}^-)\to\Lambda_c K^*\to\Lambda_c K\pi$&\multicolumn{2}{c|}{$0.02$}&&
\\ \cline{4-6}
&&&$\Xi_c({1\over2}^-)\to\Xi_c^\prime\rho\to\Xi_c^\prime\pi\pi$&\multicolumn{2}{c|}{$1.2^{+6.9}_{-1.2}$}&&
\\ \cline{4-6}
&&&$\Xi_c({1\over2}^-)\to\Sigma_c K^*\to\Sigma_c K \pi$&\multicolumn{2}{c|}{$1\times 10^{-3}$}&&
\\ \cline{4-6}
&&&$\Xi_c({1\over2}^-)\to\Xi_c^*\rho\to\Xi_c^*\pi\pi$&\multicolumn{2}{c|}{$0.09$}&&
\\ \cline{1-2} \cline{4-8}
\multirow{9}{*}{$\Xi_c(2P,{3\over2}^-)$}&\multirow{9}{*}{$3.18\pm 0.12$}&&$\Xi_c({3\over2}^-)\to\Xi_ c^{*}\pi$&$2.5^{+20.9}_{-~2.5}$&$3\times10^{-3}$&\multirow{9}{*}{$24^{+131}_{-~20}$}&\multirow{9}{*}{$\Xi_c(3123)$}
\\ \cline{4-6}
&&&$\Xi_c({3\over2}^-)\to\Sigma_c^* K$&$20^{+129}_{-~20}$&$2\times10^{-3}$&&
\\ \cline{4-6}
&&&$\Xi_c({3\over2}^-)\to\Xi_c^{\prime}\pi$&--&$0.03$&&
\\ \cline{4-6}
&&&$\Xi_c({3\over2}^-)\to\Sigma_c K$&--&$0.02$&&
\\ \cline{4-6}
&&&$\Xi_c({3\over2}^-)\to \Xi_c\rho\to\Xi_c\pi\pi$&\multicolumn{2}{c|}{$0.10^{+0.64}_{-0.10}$}&&
\\ \cline{4-6}
&&&$\Xi_c({3\over2}^-)\to \Lambda_c K^*\to\Lambda_c K \pi$&\multicolumn{2}{c|}{$0.03$}&&
\\ \cline{4-6}
&&&$\Xi_c({3\over2}^-)\to \Xi^\prime_c\rho\to\Xi^\prime_c\pi\pi$&\multicolumn{2}{c|}{$0.27^{+1.52}_{-0.27}$}&&
\\ \cline{4-6}
&&&$\Xi_c({3\over2}^-)\to \Sigma_c K^*\to\Sigma_c K\pi$&\multicolumn{2}{c|}{$2\times 10^{-3}$}&&
\\ \cline{4-6}
&&&$\Xi_c({3\over2}^-)\to \Xi^*_c\rho\to\Xi^*_c\pi\pi$&\multicolumn{2}{c|}{$0.18^{+1.01}_{-0.18}$}&&
\\ \hline \hline
\end{tabular}}
\label{tab:decayc311rho}
\end{center}
\end{table*}

\section*{Acknowledgments}

We are grateful to Shi-Lin Zhu and Er-Liang Cui for the helpful discussions.
This project is supported by
the National Natural Science Foundation of China under Grants No.~11975033, No.~12075019, and No.~12005172,
the Jiangsu Provincial Double-Innovation Program under Grant No.~JSSCRC2021488,
and
the Fundamental Research Funds for the Central Universities.

\appendix

\section{$1P$ and $2P$-wave charmed baryons from doublet $[\mathbf{\bar 3}_F, 1, 0, \lambda]$}
\label{sec:charmedsumrule}

In this appendix we study the $1P$ and $2P$-wave charmed baryons belonging to the $[\mathbf{\bar 3}_F, 1, 0, \lambda]$ doublets. We apply the QCD sum rule method to study their mass spectrum, and the obtained results are summarized in Table~\ref{tab:pwavebparameter2}. We apply the light-cone sum rule method to study their decay properties, and the obtained results are summarized in Table~\ref{tab:decayc310lambda}. These results suggest that the charmed baryons $\Lambda_c({1\over2}^-,1P/2P)$, $\Lambda_c({3\over2}^-,1P/2P)$, $\Xi_c({1\over2}^-,1P)$ and $\Xi_c({3\over2}^-,1P/2P)$ belonging to the $[\mathbf{\bar 3}_F,1,0,\lambda]$ doublet can not be easily used to explain the $\Lambda_c(2595)$, $\Lambda_c(2625)$, $\Lambda_c(2910)$, $\Lambda_c(2940)$, $\Xi_c(2790)$, $\Xi_c(2815)$, and $\Xi_c(3123)$.

\begin{table*}[hbtp]
\begin{center}
\renewcommand{\arraystretch}{1.5}
\caption{Parameters of the $1P$- and $2P$-wave charmed baryons belonging to the $[\mathbf{\bar 3}_F,1,0,\lambda]$ doublets, calculated using the method of QCD sum rules within the framework of heavy quark effective theory. Decay constants in the last column satisfy $f_{\Xi_c^+} = f_{\Xi_c^0}$.}
\begin{tabular}{ c | c | c | c | c c | c | c}
\hline\hline
\multirow{2}{*}{~~B~~} & $\omega_c$ & ~~~Working region~~~ & ~~~~~~~$\overline{\Lambda}$~~~~~~~ & ~~~Baryon~~~ & ~~~~Mass~~~~~ & ~Difference~ & Decay constant
\\                                               & (GeV) & (GeV)      & (GeV)                &                                ($j^P$)       & (GeV)      & (MeV)        & (GeV$^{4}$)
\\
\hline\hline
 \multirow{2}{*}{ $\Lambda_c(1P)$} &\multirow{2}{*}{ $1.00$ }& \multirow{2}{*}{$T=0.279$} &\multirow{2}{*}{ $0.99\pm 0.01$} & $\Lambda_c(1/2^-)$ & $2.76 \pm0.03$ &\multirow{2}{*}{ $7\pm3$}& $0.016 \pm 0.002~(\Lambda^+_c(1/2^-))$
\\ \cline{5-6}\cline{8-8}
& & & &$\Lambda_c(3/2^-)$ & $2.77\pm0.03$ & &$0.007 \pm 0.001~(\Lambda^+_c(3/2^-))$
\\ \cline{1-8}
                                                 \multirow{2}{*}{$\Xi_c(1P)$} & \multirow{2}{*}{$1.28$} & \multirow{2}{*}{$T=0.271$} & \multirow{2}{*}{$1.07 \pm 0.03$} & $\Xi_c(1/2^-)$ & $2.81\pm 0.04$ & \multirow{2}{*}{$9\pm 3$} & $0.022 \pm 0.003~(\Xi^0_c(1/2^-))$
                                                 \\ \cline{5-6}\cline{8-8}
 & & & & $\Xi_c(3/2^-)$ & $2.82 \pm 0.04$ & &$0.010 \pm 0.001~(\Xi_c^0(3/2^-))$
\\ \cline{1-8}
\multirow{2}{*}{ $\Lambda_c(2P)$} &\multirow{2}{*}{ $1.53$ }& \multirow{2}{*}{$T=0.279$} &\multirow{2}{*}{ $1.31\pm 0.07$} & $\Lambda_c(1/2^-)$ & $3.08 \pm0.12$ &\multirow{2}{*}{ $13\pm4$}& $0.026 \pm 0.007~(\Lambda^+_c(1/2^-))$
\\ \cline{5-6}\cline{8-8}
 & & & &$\Lambda_c(3/2^-)$ & $3.09 \pm0.12$ & &$0.012 \pm 0.003~(\Lambda^+_c(3/2^-))$
\\ \cline{1-8}
                                                \multirow{2}{*}{$\Xi_c(2P)$} & \multirow{2}{*}{$1.46$} & \multirow{2}{*}{$T=0.271$} & \multirow{2}{*}{$1.37 \pm 0.07$} & $\Xi_c(1/2^-)$ & $3.12\pm 0.11$ & \multirow{2}{*}{$14\pm 4$} & $0.019 \pm 0.009~(\Xi^+_c(1/2^-))$
                                                 \\ \cline{5-6}\cline{8-8}
 & & & & $\Xi_c(3/2^-)$ & $3.13 \pm0.11$ & &$0.009 \pm 0.004~(\Xi_c^+(3/2^-))$
\\ \hline\hline
\end{tabular}
\label{tab:pwavebparameter2}
\end{center}
\end{table*}

\begin{table*}[hbt]
\begin{center}
\renewcommand{\arraystretch}{1.5}
\caption{Decay properties of the $1P$ and $2P$-wave charmed baryons belonging to the $[\mathbf{\bar 3}_F, 1, 0, \lambda]$ doublets, calculated using the method of light-cone sum rules within the framework of heavy quark effective theory. }
\setlength{\tabcolsep}{0.1mm}{
\begin{tabular}{ c | c | c | c | c | c | c}
\hline\hline
Baryon & ~~~~~~Mass~~~~~~ & Difference & \multirow{2}{*}{~~~~~~~~~Decay channels~~~~~~~~~}  & ~$S$-wave width~  & ~$D$-wave width~ & ~Total width~
\\  ($j^P$) & ({GeV})& ({MeV}) & & ({MeV}) & ({MeV}) & ({MeV})
\\ \hline\hline
\multirow{3}{*}{$\Lambda_c(1P,{1\over2}^-)$}& \multirow{3}{*}{$2.76\pm 0.03$} &\multirow{6}{*}{$7\pm 3$}&$\Lambda_c({1\over2}^-)\to \Sigma_c\pi\to\Lambda_c\pi\pi$&$>500$&--&\multirow{3}{*}{$>500$}
\\ \cline{4-6}
&&&$\Lambda_c({1\over2}^-)\to\Sigma_c^*\pi\to\Lambda_c\pi\pi$&--&$2\times10^{-6}$&
\\ \cline{4-6}
&&&$\Lambda_c({1\over2}^-)\to \Lambda_c\rho\to \Lambda_c\pi\pi$& \multicolumn{2}{c|}{$1\times10^{-3}$ }&
\\ \cline{1-2} \cline{4-7}
\multirow{3}{*}{$\Lambda_c(1P,{3\over2}^-)$}&\multirow{3}{*}{$2.77\pm 0.03$}&&$\Lambda_c({3\over2}^-)\to \Sigma_c \pi$ & -- & $0.06$&\multirow{3}{*}{$29^{+41}_{-21}$}
\\ \cline{4-6}
&&&$\Lambda_c({3\over2}^-)\to\Sigma_c^*\pi\to\Lambda_c\pi\pi$&$29^{+41}_{-21}$&$3\times10^{-5}$&
\\ \cline{4-6}
&&&$\Lambda_c({3\over2}^-)\to \Lambda_c\rho\to \Lambda_c\pi\pi$&\multicolumn{2}{c|}{$0.01$}&
\\ \cline{1-7}
\multirow{5}{*}{$\Lambda_c(2P,{1\over2}^-)$}& \multirow{5}{*}{$3.08\pm 0.12$} &\multirow{10}{*}{$13\pm 4$}&$\Lambda_c({1\over2}^-)\to \Sigma_c\pi$&$>2000$&--&\multirow{5}{*}{$>2000$}
\\ \cline{4-6}
&&&$\Lambda_c({1\over2}^-)\to\Sigma_c^*\pi$&--&$0.92^{+1.77}_{-0.74}$&
\\ \cline{4-6}
&&&$\Lambda_c({1\over2}^-)\to \Lambda_c\rho\to \Lambda_c\pi\pi$& \multicolumn{2}{c|}{$1.8^{+2.7}_{-1.3}$ }&
\\ \cline{4-6}
&&&$\Lambda_c({1\over2}^-)\to \Sigma_c\rho\to \Sigma_c\pi\pi$& \multicolumn{2}{c|}{$3\times10^{-3}$ }&
\\ \cline{4-6}
&&&$\Lambda_c({1\over2}^-)\to \Sigma^*_c\rho\to \Sigma^*_c\pi\pi$& \multicolumn{2}{c|}{$1\times10^{-4}$}&
\\ \cline{1-2}\cline{4-7}
\multirow{5}{*}{$\Lambda_c(2P,{3\over2}^-)$}&\multirow{5}{*}{$3.09\pm 0.12$}&&$\Lambda_c({3\over2}^-)\to \Sigma_c \pi$ & -- & $1.5^{+2.8}_{-1.2}$&\multirow{5}{*}{$>500$}
\\ \cline{4-6}
&&&$\Lambda_c({3\over2}^-)\to\Sigma_c^*\pi$&$>500$&$0.19^{+0.36}_{-0.15}$&
\\ \cline{4-6}
&&&$\Lambda_c({3\over2}^-)\to \Lambda_c\rho\to \Lambda_c\pi\pi$&\multicolumn{2}{c|}{$1.7^{+2.6}_{-1.2}$}&
\\ \cline{4-6}
&&&$\Lambda_c({3\over2}^-)\to \Sigma^*_c\rho\to \Sigma^*_c\pi\pi$&\multicolumn{2}{c|}{$8\times10^{-5}$}&
\\ \hline\hline
\multirow{3}{*}{$\Xi_c(1P,{1\over2}^-)$}&\multirow{3}{*}{$2.81\pm 0.04$}&\multirow{6}{*}{$9\pm 3$}&$\Xi_c({1\over2}^-)\to \Xi_c^{\prime}\pi$&$>1800$&--&\multirow{3}{*}{$>1800$}
\\ \cline{4-6}
&&&$\Xi_c({1\over2}^-)\to\Xi_c^*\pi\to\Xi_c\pi\pi$&--&$2\times10^{-4}$&
\\ \cline{4-6}
&&&$\Xi_c({1\over2}^-)\to\Xi_c\rho\to\Xi_c\pi\pi$&\multicolumn{2}{c|}{$6\times10^{-3}$}&
\\ \cline{1-2} \cline{4-7}
\multirow{3}{*}{$\Xi_c(1P,{3\over2}^-)$}&\multirow{3}{*}{$2.82\pm 0.04$}&&$\Xi_c({3\over2}^-)\to\Xi_ c^{*}\pi\to\Xi_c\pi\pi$&$>300$&$1\times10^{-3}$&\multirow{3}{*}{$>300$}
\\ \cline{4-6}
&&&$\Xi_c({3\over2}^-)\to\Xi_c^{\prime}\pi$&--&$0.23^{+0.35}_{-0.17}$&
\\ \cline{4-6}
&&&$\Xi_c({3\over2}^-)\to \Xi_c\rho\to\Xi_c\pi\pi$&\multicolumn{2}{c|}{$0.02$}&
\\ \hline
\multirow{8}{*}{$\Xi_c(2P,{1\over2}^-)$}&\multirow{8}{*}{$3.12\pm 0.11$}&\multirow{17}{*}{$14\pm 4$}&$\Xi_c({1\over2}^-)\to \Xi_c^{\prime}\pi$&$900^{+3700}_{-~900}$&--&\multirow{8}{*}{$2200^{+7100}_{-1600}$}
\\ \cline{4-6}
&&&$\Xi_c({1\over2}^-)\to\Sigma_c K$&$1300^{+6100}_{-1300}$&--&
\\ \cline{4-6}
&&&$\Xi_c({1\over2}^-)\to\Xi_c^*\pi$&--&$0.12^{+0.51}_{-0.12}$&
\\ \cline{4-6}
&&&$\Xi_c({1\over2}^-)\to\Sigma_c^* K$&--&$0.03$&
\\ \cline{4-6}
&&&$\Xi_c({1\over2}^-)\to\Xi_c\rho\to\Xi_c\pi\pi$&\multicolumn{2}{c|}{$4.7^{+14.0}_{-~3.8}$}&
\\ \cline{4-6}
&&&$\Xi_c({1\over2}^-)\to\Lambda_c K^*\to\Lambda_c K\pi$&\multicolumn{2}{c|}{$0.19^{+0.77}_{-0.19}$}&
\\ \cline{4-6}
&&&$\Xi_c({1\over2}^-)\to\Xi_c^\prime\rho\to\Xi_c^\prime\pi\pi$&\multicolumn{2}{c|}{$6\times10^{-4}$}&
\\ \cline{4-6}
&&&$\Xi_c({1\over2}^-)\to\Sigma_c K^*\to\Sigma_c K \pi$&\multicolumn{2}{c|}{$6\times10^{-6}$}&
\\ \cline{4-6}
&&&$\Xi_c({1\over2}^-)\to\Xi_c^*\rho\to\Xi_c^*\pi\pi$&\multicolumn{2}{c|}{$4\times10^{-5}$}&
\\ \cline{1-2} \cline{4-7}
\multirow{9}{*}{$\Xi_c(2P,{3\over2}^-)$}&\multirow{9}{*}{$3.13\pm 0.11$}&&$\Xi_c({3\over2}^-)\to\Xi_ c^{*}\pi$&$200^{+700}_{-200}$&$0.02$&\multirow{9}{*}{$500^{+1200}_{-~400}$}
\\ \cline{4-6}
&&&$\Xi_c({3\over2}^-)\to\Sigma_c^* K$&$300^{+1000}_{-~300}$&$0.01$&
\\ \cline{4-6}
&&&$\Xi_c({3\over2}^-)\to\Xi_c^{\prime}\pi$&--&$0.15^{+0.62}_{-0.15}$&
\\ \cline{4-6}
&&&$\Xi_c({3\over2}^-)\to\Sigma_c K$&--&$0.17^{+0.72}_{-0.17}$&
\\ \cline{4-6}
&&&$\Xi_c({3\over2}^-)\to \Xi_c\rho\to\Xi_c\pi\pi$&\multicolumn{2}{c|}{$0.42^{+1.64}_{-0.42}$}&
\\ \cline{4-6}
&&&$\Xi_c({3\over2}^-)\to \Lambda_c K^*\to\Lambda_c K \pi$&\multicolumn{2}{c|}{$0.20^{+0.78}_{-0.19}$}&
\\ \cline{4-6}
&&&$\Xi_c({3\over2}^-)\to \Xi^*_c\rho\to\Xi^*_c\pi\pi$&\multicolumn{2}{c|}{$1\times10^{-5}$}&
\\ \hline \hline
\end{tabular}}
\label{tab:decayc310lambda}
\end{center}
\end{table*}

%


\end{document}